\documentclass{aa}  

\usepackage{graphicx}
\usepackage{txfonts}
\usepackage{multirow}
\usepackage{colortbl}

\usepackage{hyperref}  
\hypersetup{colorlinks=true,linkcolor=[rgb]{1.,0.2,0.2},citecolor=[rgb]{0.1,0.4,1.},filecolor=[rgb]{0.7,0.2,0.2},urlcolor=[rgb]{0.7,0.2,0.2}}

\usepackage{color}
\definecolor{blue}{rgb}{0., 0., 1}
\definecolor{lightblue}{rgb}{0.1,0.4,1.}

\newcommand{\ci}[1]{{\color{lightblue}{#1}}}
\newcommand {\CL}{M0416}
\newcommand {\LT}{\texttt{LensTool}}
\newcommand {\T}{Table\,}
\newcommand {\Hubble}{{\it Hubble}}
\newcommand {\Sec}{Sect.\,}
\newcommand {\Fig}{Fig.\,}
\newcommand {\Eq}{Eq.\,}

\newcommand {\SLOT}{\texttt{SLOT}}

\defcitealias{Bergamini_2021}{B21}
\defcitealias{Richard_2021}{R21}

\begin{document} 

\title{A state-of-the-art strong lensing model of MACS~J0416.1$-$2403 with the largest sample of spectroscopic multiple images}

\author{
P.~Bergamini \inst{\ref{unimi}, \ref{inafbo}} \fnmsep\thanks{E-mail: \href{mailto:pietro.bergamini@inaf.it}{pietro.bergamini@unimi.it}} \and
C.~Grillo \inst{\ref{unimi},\ref{inafmilano}} \and
P.~Rosati \inst{\ref{unife},\ref{inafbo}} \and
E.~Vanzella \inst{\ref{inafbo}} \and
U.~Me{\v{s}}tri{\'c} \inst{\ref{inafbo}} \and
A.~Mercurio \inst{\ref{inafna}} \and
A.~Acebron \inst{\ref{unimi},\ref{inafmilano}} \and
G.~B.~Caminha \inst{\ref{max_plank}} \and
G.~Granata \inst{\ref{unimi},\ref{inafmilano}} \and
M.~Meneghetti \inst{\ref{inafbo}} \and
G.~Angora \inst{\ref{unife},\ref{inafna}} \and
M.~Nonino \inst{\ref{inafts}} 
}
\institute{
Dipartimento di Fisica, Universit\`a  degli Studi di Milano, via Celoria 16, I-20133 Milano, Italy \label{unimi}
\and
INAF -- OAS, Osservatorio di Astrofisica e Scienza dello Spazio di Bologna, via Gobetti 93/3, I-40129 Bologna, Italy \label{inafbo}
\and
INAF - IASF Milano, via A. Corti 12, I-20133 Milano, Italy
\label{inafmilano}
\and
Dipartimento di Fisica e Scienze della Terra, Universit\`a degli Studi di Ferrara, via Saragat 1, I-44122 Ferrara, Italy \label{unife}
\and
INAF -- Osservatorio Astronomico di Capodimonte, Via Moiariello 16, I-80131 Napoli, Italy \label{inafna}
\and
Max-Planck-Institut f\"ur Astrophysik, Karl-Schwarzschild-Str. 1, D-85748 Garching, Germany \label{max_plank}
\and
INAF -- Osservatorio Astronomico di Trieste, via G. B. Tiepolo 11, I-34143, Trieste, Italy \label{inafts}
           }

   \date{Received February 14, 2020; accepted February 14, 2020}

  \abstract
  {
    The combination of multi-band imaging from the Hubble Space Telescope with Multi-Unit Spectroscopic Explorer integral field spectroscopy, obtained at the Very Large Telescope, has recently driven remarkable progress in strong lensing (SL) modeling of galaxy clusters. From a few tens of multiple images with photometric redshifts per cluster, a new generation of high-precision SL models have recently been developed, by exploiting in some cases over a hundred of spectroscopically confirmed multiple images and cluster member galaxies. A further step forward is expected with James Webb Space Telescope observations of SL clusters (from hundreds to possibly a thousand of multiple images).
    In this context, we present a new, state-of-the-art SL model of the galaxy cluster MACS J0416.1$-$2403, utilizing 237 spectroscopically confirmed multiple images, which is the largest sample of secure multiply lensed sources utilized to date. In addition, this model incorporates stellar kinematics information of 64 cluster galaxies and the hot-gas mass distribution of the cluster, determined from Chandra X-ray observations. The observed positions of the many multiple images are reproduced with a remarkable average accuracy of $0.43\arcsec$. To further assess the reliability of this lens model and to highlight the improvement over previously published models, we show the extended surface brightness reconstruction of several lensed galaxies through a newly developed forward modeling software. The comparison with other SL models of the same cluster demonstrates that this new model is better suited to accurately reproduce the positions, shapes and fluxes of the observed multiple images. Besides a robust characterization of the total mass distribution of the cluster, our model can provide accurate and precise magnification maps that are key to studying the intrinsic physical properties of faint, high-redshift lensed sources. 
    The model is made publicly available through our newly developed Strong Lensing Online Tool (or \texttt{SLOT}), that thank to a simple graphical interface allows astronomers (including lensing non-experts) to take full advantage of the predictive power of the model, including statistical uncertainties on the relevant quantities associated to the multiply lensed sources.
  }

   \keywords{Galaxies: clusters: general -- Gravitational lensing: strong -- cosmology: observations -- dark matter -- galaxies: kinematics
and dynamics
            }
   \titlerunning{A state-of-the-art strong lensing model of MACS~J0416.1$-$2403}
   \authorrunning{P.~Bergamini et al.}
   \maketitle


\section{Introduction}

\begin{figure*}
	\centering
	\includegraphics[width=1\linewidth]{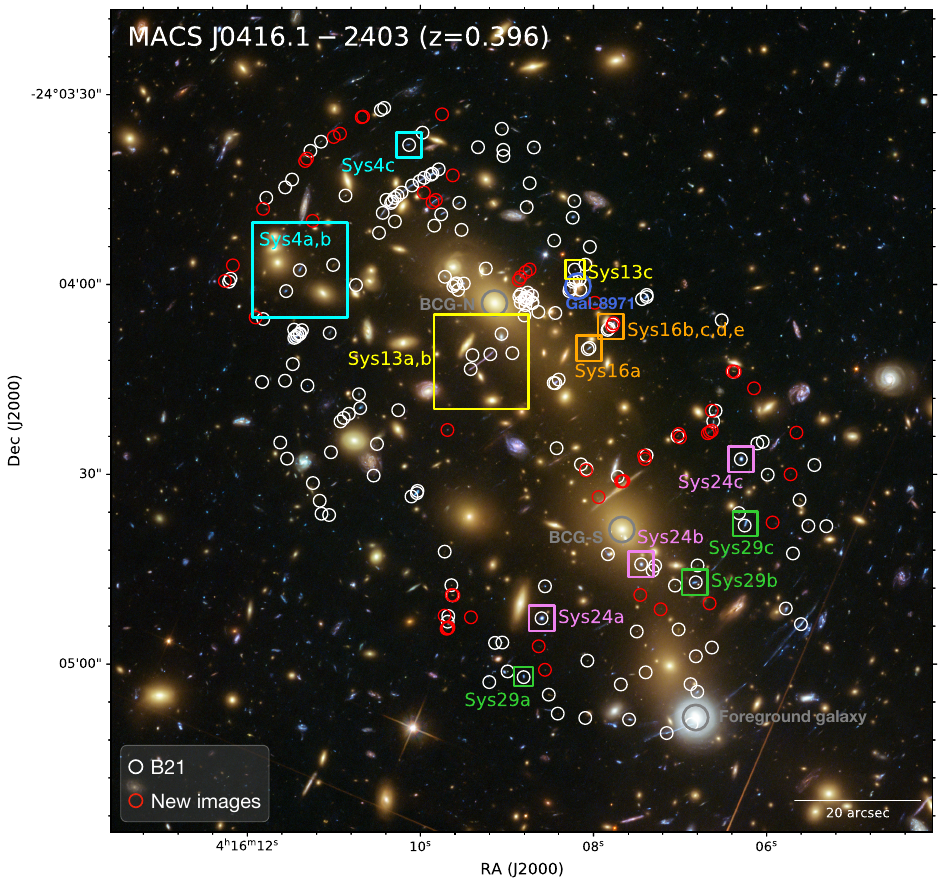}
	\caption{Color-composite image (credits \href{https://esahubble.org/images/heic1416a/}{NASA/ESA}) of the galaxy cluster \CL. White circles show the sample of 182 multiple images in common with the lens model by \cite{Bergamini_2021}. The new 55 images added to the previous ones and used to optimize the lens model described in this work are plotted in red. A blue circle marks the galaxy optimized separately from the cluster member scaling relations (see \Sec\ref{sec:model_description}). Colored squares highlight the systems of multiple images analyzed using a new forward modeling approach (described in \Sec\ref{sec:results}) that provides a further validation of our lens model. The two BCGs (BCG-N and BCG-S) and a foreground galaxy at $z=0.112$ are encircled in gray.}
	\label{fig:RGB}
\end{figure*}

Strong gravitational lensing has become a powerful technique to characterize the total mass distribution in the core of galaxy clusters \citep[see e.g.,][]{Caminha_macs1206, Diego2020, Sharon2020, Pignataro2021, Jauzac2021}, to discover and study the physical properties of high-redshift galaxies \citep{Coe_2013, Bouwens_2014, Zitrin2015, Hashimoto_2018}, even resolving their structures at sub-kpc scales \citep{johnson17, Dessauges2017, Mestric2022}, and to investigate the expansion and the geometry of the Universe \citep{Jullo2010, Caminha_rxc2248, Grillo_2018, Grillo_2020, Caminha2022}.
All of these studies rely on accurate strong lensing models, where the accuracy of the latter critically depends on the number of spectroscopically confirmed multiple images \citep{Grillo_2015, Johnson2016, Caminha_2019}.

The recent surge of high-quality photometric and spectroscopic data on galaxy clusters has allowed for some important progress in cluster lens models and their subsequent applications. In particular, the combination of high-resolution, panchromatic and deep imaging data from the \textit{Hubble Space Telescope} (\Hubble) with follow-up spectroscopy with the Multi-Unit Spectroscopic Explorer \citep[MUSE,][]{Bacon_MUSE} on the Very Large Telescope (VLT) has lead to a significant increase in the number of multiple images securely identified in the core of galaxy clusters \citep[see e.g.,][hereafter \citetalias{Richard_2021}]{Grillo2016, Karman_2017, Lagattuta_2017, Bergamini2022, Richard_2021}. 

MACS J0416.1$-$2403 (hereafter \CL, see \Fig\ref{fig:RGB}), discovered within the Massive Cluster Survey \citep{Ebeling2001}, has been the target of several imaging programs with \Hubble\, such as the Cluster Lensing And Supernova survey with Hubble \citep[CLASH,][]{Postman_2012_clash}, the Hubble Frontier Fields \citep[HFF,][]{Lotz_2017HFF} and the Beyond Ultra-deep Frontier Fields And Legacy Observations \citep[BUFFALO,][]{Steinhardt2020}.
M0416 is a massive, as found from weak-lensing studies \citep{Umetsu_2014}\footnote{With a mass estimate of $M_{200c}=(1.04\pm0.22)\times 10^{15} \rm \, M_\odot$.}, and X-ray luminous \citep{Mann2012} galaxy cluster at redshift $z = 0.396$.
The system presents a complex and mostly bi-modal mass distribution, which is likely the result of a pre-collisional phase \citep{Balestra_2016}.
In addition, its highly elongated geometry, typical of merging clusters, makes \CL\ a remarkably efficient gravitational lens compared to other cluster lenses, as illustrated by the first strong lensing analysis \citep{Zitrin_2013}.
Taking advantage of the CLASH imaging data and the CLASH-VLT spectroscopic follow-up program \citep{Balestra_2016}, \citet{Grillo_2015} presented a strong lensing model of \CL\ including 30 spectroscopic multiple images from 10 different sources.
Subsequent parametric and free-form studies were carried out, combining weak and strong lensing analyses and testing the impact of line-of-sight mass structures \citep{Jauzac_2014, Richard_2014, Jauzac_2015, Hoag_2016, Chirivi_2018}. In particular, \citet{Hoag_2016} included spectroscopic data from the Grism Lens-Amplified Survey from Space \citep[GLASS,][]{Treu_2015}, resulting in 30 secure multiple images belonging to 15 distinct sources.
A high-precision strong lens model exploiting the first MUSE observations of \CL, that lead to the identification of a large sample of 102 spectroscopic multiple images from 37 background sources, was then presented in \citet{Caminha_macs0416}.
This model was improved upon with the inclusion of the mass component associated to the hot gas \citep{Bonamigo_2017, Bonamigo_2018}, and the kinematic measurements of a large sample of clusters galaxies, that independently constrained the sub-halo mass component \citep{Bergamini_2019}.
In \citet[][hereafter \citetalias{Bergamini_2021}]{Bergamini_2021} we combined the analysis of the MUSE Deep Lens Field \citep[MDLF, see][]{Vanzella_2020} carried out in the northeast region of the cluster, with a careful re-inspection of the \Hubble\ images, and identified 182 secure images from 66 different background sources or source substructures, $\sim 80\%$ more multiple images compared to previous works \citep{Caminha_macs0416, Bergamini_2019}.
Thanks to the large number of observational constraints and the detailed modeling of the cluster total mass distribution, the \citetalias{Bergamini_2021} lens model achieved a high level of precision and accuracy. This was shown by the reproduction of the relative distances and orientations of pairs of multiply imaged clumps, inside well resolved sources in the vicinity of the critical lines, with typical values of less than $0.33\arcsec$ and $5.9^{\circ}$, respectively, for 90\% of the image pairs. 

\begin{figure}
	\centering
	\includegraphics[width=1\linewidth]{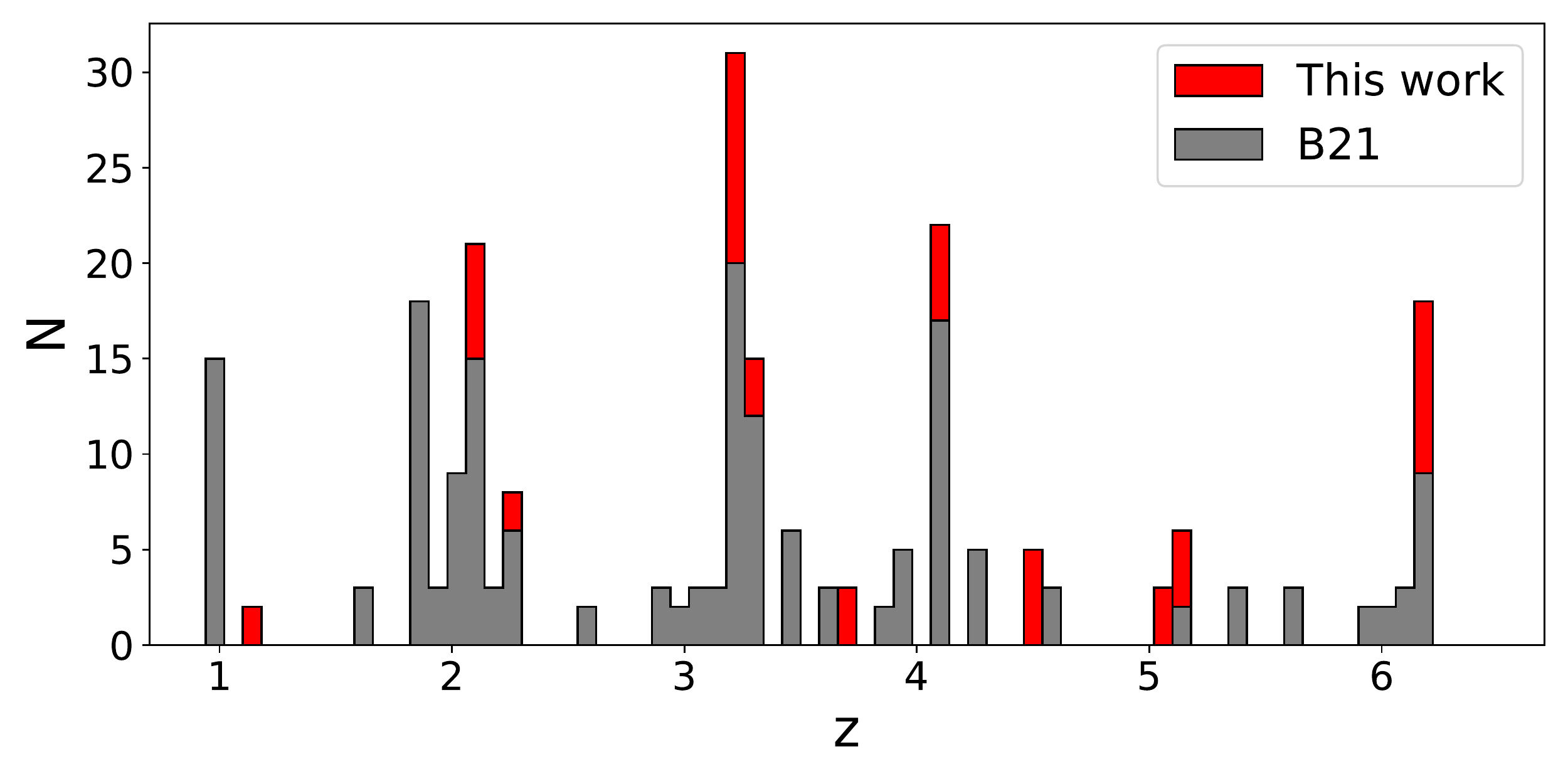}
	\caption{Redshift distribution of the observed 237 multiple images exploited to constrain the reference lens model described in this work. The multiple images used by \citetalias{Bergamini_2021} are plotted in gray, while the new images are shown in red.}
	\label{fig:hist}
\end{figure}

\begin{figure}
	\centering
	\includegraphics[width=1\linewidth]{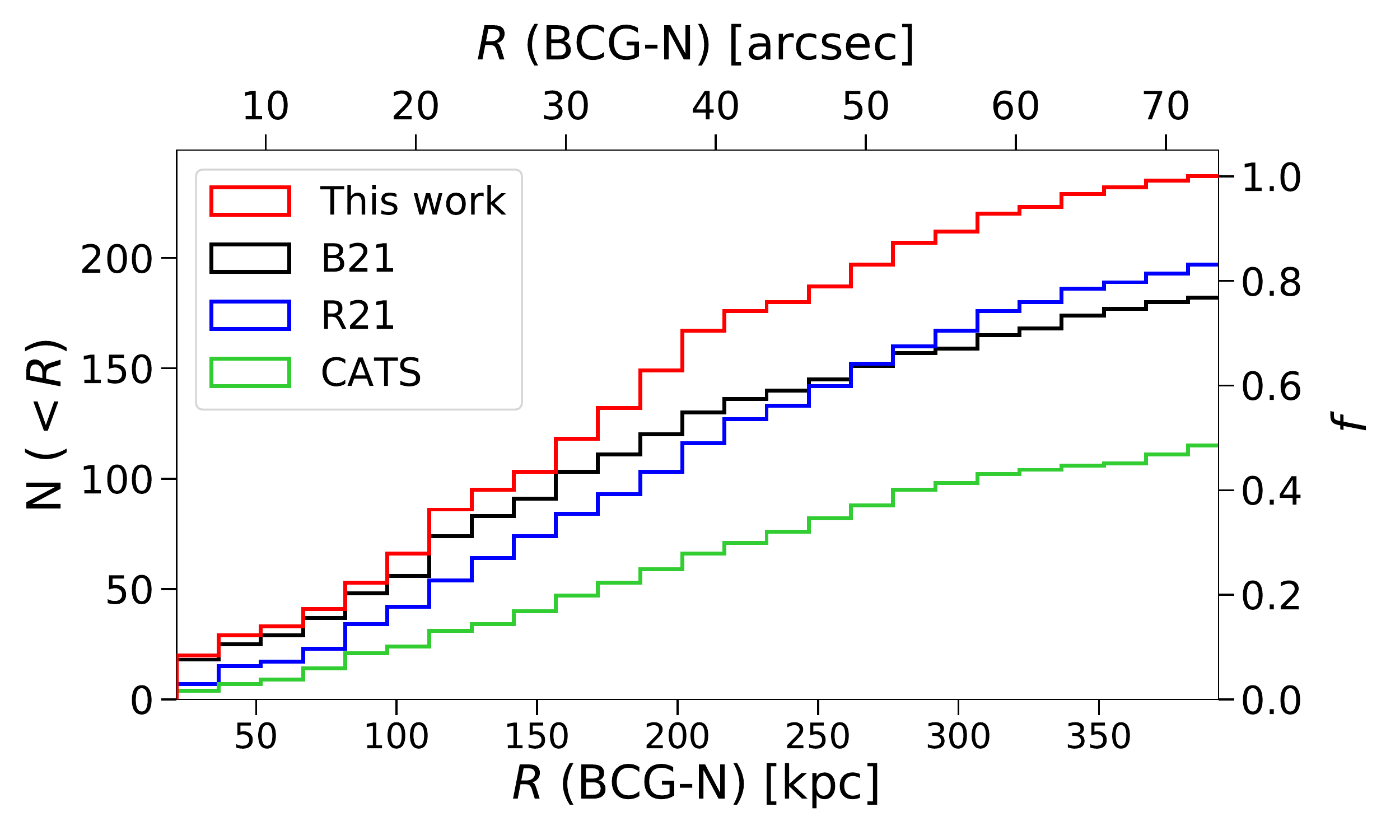}
	\caption{Cumulative distributions of the number of multiple images as a function of their projected distance from the northern BCG (BCG-N) of \CL. We plot in red the distribution of the images used as constraints in the new model described in this work (237 multiple images in total). As a comparison, we show in black, blue, and green the multiple images from the \citetalias{Bergamini_2021}, \citetalias{Richard_2021}, and CATS models, respectively.}
	\label{fig:image_distribution}
\end{figure}

In this work, we further exploit the \Hubble\ multi-color imaging and MUSE spectroscopy to develop a refined, high-precision strong lensing model of \CL. We have identified 55 additional spectroscopically confirmed multiple images, an increase of $\sim 30\%$ compared to the previous catalog published in \citetalias{Bergamini_2021}. The new set of multiple images mainly consists of multiply lensed clumps within resolved extended sources, which are particularly useful to constrain locally the position of the critical lines \citep[as shown in][]{Grillo2016, Bergamini_2021, Bergamini2022}.
The final multiple image sample consists of 237 multiple images from 88 different background sources, spanning a wide redshift range $z=$ 0.94 - 6.63 (see \Fig\ref{fig:hist}).
Thus, this work presents  the \textit{largest set of secure multiple images ever constructed} (see \Fig\ref{fig:image_distribution}), paving the way for a new generation of cluster strong lensing models, with an unprecedented level of accuracy and precision.

The paper is organized as follows. In \Sec\ref{sec:Data}, we present the imaging and spectroscopic data used to develop the lens model. \Sec\ref{sec:model_description} describes the sample of multiple images used as model constraints and the adopted mass parametrization. In \Sec\ref{sec:results}, we discuss the results of the mass model, and compare it to that presented in \citetalias{Bergamini_2021}.
The new lens model of \CL, including the largest set of secure multiple image to date, is then used to reconstruct the shape and luminosity of several multiply-imaged sources, highlighting the high precision and accuracy achieved. 
The main conclusions of this work are summarized in \Sec\ref{sec:conclusions}.

Throughout this work, we adopt a flat $\Lambda$CDM cosmology
with $\Omega_{\rm m} = 0.3$ and $H_0= 70\,\mathrm{km\,s^{-1}\,Mpc^{-1}}$. Using this cosmology, a projected distance of $1\arcsec$ corresponds to a physical scale of 5.34 kpc at the \CL\ redshift of $z=0.396$. All magnitudes are given in the AB system.


\section{Data}
\label{sec:Data}
The updated, high-precision strong lensing model for the galaxy cluster \CL\ presented in this work is based on the same observational dataset described in \citetalias{Bergamini_2021} and briefly summarized hereafter:

\begin{enumerate}
    \item {\it Photometric data:} \Hubble\ multi-band observations were collected within the CLASH survey (16 filters) and the HFF program (7 filters).
    
    \item {\it Spectroscopic data:} VLT/VIMOS observations provided redshift measurements over a $\sim\! 20\arcmin$ field of view \citep[see][]{Balestra_2016}. Several MUSE observations were also performed on the cluster core. In particular, one MUSE pointing (GTO 094.A-0115B, P.I. J. Richard) was centered on the northeast (NE) region of \CL\ (2h of exposure and $0.6\arcsec$ seeing). A second MUSE observation (094.A0525(A), P.I. F. E. Bauer) was pointed to the southwest region of the cluster (11h of integration and $1.0\arcsec$ seeing). We refer to the work by \cite{Caminha_macs0416} for a comprehensive description of these MUSE observations. Finally, an ultra-deep MUSE observation on the NE region of \CL\ was performed through the observational program 0100.A-0763(A) with P.I. E. Vanzella \citep{Vanzella_2020}. This is the deepest MUSE observation obtained on a galaxy cluster to date, reaching an integration time of 17.1h in most of the NE field of \CL\, with a seeing of approximately $0.6\arcsec$.
\end{enumerate}

\section{Model description}
\label{sec:model_description}
The \CL\ lens model presented in this work has been developed using the publicly available software \LT\footnote{\url{https://projets.lam.fr/projects/lenstool/wiki}}\ \citep{Kneib_lenstool, Jullo_lenstool, Jullo_Kneib_lenstool}. To determine the total mass distribution of a galaxy cluster, \LT\ exploits a Bayesian approach that minimizes the following $\chi^2$ function, which quantifies how good is the lens model at reproducing the point-like positions of the 237 observed multiple images included in the sample:

\begin{equation}
    \label{eq.: chi_lt}
    \chi^2(\pmb{\xi}) := \sum_{j=1}^{N_{fam}} \sum_{i=1}^{N_{im}^j} \left(\frac{\left\| \mathbf{x}_{i,j}^{pred}(\pmb{\xi}) - \mathbf{x}_{i,j}^{obs} \right\|}{\Delta x_{i,j}}\right)^2.
\end{equation}

\noindent In this equation, $\mathbf{x}_{i,j}^{obs}$ is the observed position of the $i^{th}$ multiple image of the $j^{th}$ background source (images from the same source are called a family of multiple images), $\mathbf{x}_{i,j}^{pred}$ is its predicted position, given the set of model free parameters $\pmb{\xi}$, and $\Delta x_{i,j}$ represents the positional uncertainty of the image.

Throughout this work, we quote the optimized parameter values, and their associated errors, from the $50^{th}$, $16^{th}$, and $84^{th}$ percentiles of the marginalized posterior distribution of each parameter. We note that we re-scale the $\Delta x_{i,j}$ values in such a way that the $\chi^2$ value is equal to the number of degrees of freedom ($\mathrm{dof}=2\times [N^{tot}_{im} - N_{fam}] - N_{freepar}$) of the model, before sampling the posterior distributions.

Other than the $\chi^2$, the most common figure of merit to quantify the goodness of a lens model is the root-mean-square separation between the observed and model-predicted positions of the multiple images. This is defined as: 

\begin{equation}
    \label{eq.: rms_lt}
    \Delta_{rms}=\sqrt{\frac{1}{N_{im}^{tot}}\sum_{i=1}^{N_{im}^{tot}}\left\|\boldsymbol{\Delta}_i \right\|^2} ,
\end{equation}

\noindent where $\boldsymbol{\Delta}_i=\mathbf{x}_{i}^{pred} - \mathbf{x}_{i}^{obs}$ is the separation between the model-predicted and observed positions of the $i$-th image.

\begin{table*}[]     
	\tiny
	\def\arraystretch{2.3}
	\centering          
	\begin{tabular}{|c|c|c|c|c|c|c|c|c|}
	    \cline{3-9}
		\multicolumn{2}{c|}{} & \multicolumn{7}{c|}{ \textbf{Input parameter values and assumed priors}} \\
		\cline{3-9}
		  \multicolumn{2}{c|}{} & \boldmath{$x\, \mathrm{[arcsec]}$} & \boldmath{$y\, \mathrm{[arcsec]}$} & \boldmath{$e$} & \boldmath{$\theta\ [^{\circ}]$} & \boldmath{$\sigma_{LT}\, \mathrm{[km\ s^{-1}]}$} & \boldmath{$r_{core}\, \mathrm{[arcsec]}$} & \boldmath{$r_{cut}\, \mathrm{[arcsec]}$} \\ 
          \hline

		  \multirow{8}{*}{\rotatebox[origin=c]{90}{\textbf{Cluster-scale halos}}} 
		  
		  & \boldmath{$1^{st}$} \bf{Cluster Halo} & $-15.0\,\div\,15.0$ & $-15.0\,\div\,15.0$ & $0.20\,\div\,0.90$ & $100.0\,\div\,180.0$ & $350\,\div\,1000$ & $0.0\,\div\,20.0$ & $2000.0$ \\
		  
		  & \boldmath{$2^{nd}$} \bf{Cluster Halo} & $15.0\,\div\,30.0$ & $-45.0\,\div\,-30.0$ & $0.20\,\div\,0.90$ & $90.0\,\div\,170.0$ & $350\,\div\,1200$ & $0.0\,\div\,25.0$ & $2000.0$  \\
		  
		  & \boldmath{$3^{rd}$} \bf{Cluster Halo} & $-55.0\,\div\,-25.0$ & $0.0\,\div\,30.0$ & $0.00$ & $0.0$ & $50\,\div\,750$ & $0.0\,\div\,35.0$ & $2000.0$  \\
		  
		  & \boldmath{$4^{th}$} \bf{Cluster Halo} & $-10.0\,\div\,50.0$ & $-75.0\,\div\,-15.0$ & $0.20\,\div\,0.90$ & $0.0\,\div\,180.0$ & $100\,\div\,1000$ & $0.0\,\div\,20.0$ & $2000.0$ \\
		  \cline{2-9}

		  & \boldmath{$1^{st}$} \bf{Gas Halo} & $-18.1$ & $-12.1$ & $0.12$ & $-156.8$ & $433$ & $149.2$ & $149.8$  \\
		  
		  & \boldmath{$2^{nd}$} \bf{Gas Halo} & $30.8$ & $-48.7$ & $0.42$ & $-71.5$ & $249$ & $34.8$ & $165.8$  \\
		  
		  & \boldmath{$3^{rd}$} \bf{Gas Halo} & $-2.4$ & $-1.3$ & $0.42$ & $-54.7$ & $102$ & $8.3$ & $37.6$  \\
		  
		  & \boldmath{$4^{th}$} \bf{Gas Halo} & $-20.1$ & $14.7$ & $0.40$ & $-49.3$ & $282$ & $51.7$ & $52.3$  \\

          \hline
          \multicolumn{1}{c}{}
          \\[-5ex]

          \hline

		  \multirow{3}{*}{\rotatebox[origin=c]{90}{\textbf{Subhalos}}}

		  & \bf{Gal-8971} & $13.3$ & $2.6$ & $0.00\,\div\,0.60$ & $-90.0\,\div\,90.0$ & $60\,\div\,200$ & $0.0001$ & $0.0\,\div\,50.0$ \\
		  
		  \cline{2-9}
		  
		  & \bf{Foreground gal.} & $32.0$ & $-65.6$ & $0.00$ & $0.0$ & $50\,\div\,350$ & $0.0001$ & $5.0\,\div\,100.0$ \\
		  
		  \cline{2-9}
		  
		  & \bf{Scaling relations} & $\boldsymbol{N_{gal}=}212$
		  & $\boldsymbol{m_{\mathrm{F160W}}^{ref}=}17.02$
		  & $\boldsymbol{\alpha=}0.30$ & $\boldsymbol{\sigma_{LT}^{ref}=}[248\,,\,28]$ & $\boldsymbol{\beta_{cut}=}0.60$ & $\boldsymbol{r_{cut}^{ref}=}1.0\,\div\,50.0$ & $\boldsymbol{\gamma=}0.20$\\
		  
		  \hline
		 
	\end{tabular}
	\\[6ex]
	\begin{tabular}{|c|c|c|c|c|c|c|c|c|}
	    \cline{3-9}
		\multicolumn{2}{c|}{} & \multicolumn{7}{c|}{ \textbf{Optimized output parameters}} \\
		\cline{3-9}
		  \multicolumn{2}{c|}{} & \boldmath{$x\, \mathrm{[arcsec]}$} & \boldmath{$y\, \mathrm{[arcsec]}$} & \boldmath{$e$} & \boldmath{$\theta\ [^{\circ}]$} & \boldmath{$\sigma_{LT}\, \mathrm{[km\ s^{-1}]}$} & \boldmath{$r_{core}\, \mathrm{[arcsec]}$} & \boldmath{$r_{cut}\, \mathrm{[arcsec]}$} \\ 
          \hline

		  \multirow{4}{*}{\rotatebox[origin=c]{90}{\textbf{Cluster-scale halos}}} 
		  
		  & \boldmath{$1^{st}$} \bf{Cluster Halo} & $-0.2_{-0.3}^{+0.3}$ & $0.0_{-0.3}^{+0.3}$ & $0.81_{-0.01}^{+0.01}$ & $143.9_{-0.6}^{+0.6}$ & $596_{-15}^{+13}$ & $7.3_{-0.3}^{+0.3}$ & $2000.0$ \\
		  
		  & \boldmath{$2^{nd}$} \bf{Cluster Halo} & $23.7_{-0.8}^{+0.9}$ & $-35.3_{-1.3}^{+0.8}$ & $0.88_{-0.03}^{+0.01}$ & $135.0_{-2.7}^{+1.6}$ & $480_{-59}^{+99}$ & $6.5_{-1.8}^{+2.6}$ & $2000.0$  \\
		  
		  & \boldmath{$3^{rd}$} \bf{Cluster Halo} & $-32.1_{-0.7}^{+0.6}$ & $8.8_{-0.6}^{+0.7}$ & $0.0$ & $0.0$ & $334_{-24}^{+25}$ & $8.1_{-1.3}^{+1.3}$ & $2000.0$  \\
		  
		  & \boldmath{$4^{th}$} \bf{Cluster Halo} & $21.8_{-1.2}^{+0.8}$ & $-46.7_{-1.5}^{+1.3}$ & $0.76_{-0.04}^{+0.02}$ & $122.2_{-1.9}^{+1.2}$ & $702_{-88}^{+39}$ & $13.2_{-1.1}^{+0.8}$ & $2000.0$ \\
		  \cline{2-9}
		  
          \hline
          \multicolumn{1}{c}{}
          \\[-5ex]
          \hline
		  \multirow{3}{*}{\rotatebox[origin=c]{90}{\textbf{Subhalos}}} 

		  & \bf{Gal-8971} & $13.3$ & $2.6$ & $0.52_{-0.11}^{+0.06}$ & $-40.1_{-14.7}^{+19.6}$ & $109_{-5}^{+6}$ & $0.0001$ & $18.6_{-8.3}^{+9.5}$ \\
		  
		  \cline{2-9}
		  
		  & \bf{Foreground. gal.} & $32.0$ & $-65.6$ & $0.0$ & $0.0$ & $103_{-30}^{+32}$ & $0.0001$ & $53.0_{-29.9}^{+29.8}$ \\
		  
		  \cline{2-9}
		  
		  & \bf{Scaling relations} & $\boldsymbol{N_{gal}=}212$
		  & $\boldsymbol{m_{\mathrm{F160W}}^{ref}=}17.02$
		  & $\boldsymbol{\alpha=}0.30$ & $\boldsymbol{\sigma_{LT}^{ref}=}230_{-16}^{+10}$ & $\boldsymbol{\beta_{cut}=}0.60$ & $\boldsymbol{r_{cut}^{ref}=}10.1_{-1.8}^{+2.1}$ & $\boldsymbol{\gamma=}0.20$\\
		 
		 \hline
		 
	\end{tabular}
	\\[6ex]

    \caption{Input and output optimized parameters of the \CL\ lens model presented in this work.
    {\it Top:} Input parameter values of the reference model for the galaxy cluster \CL\ presented in this work. A single number is quoted for a fixed parameter value. When a flat prior on a free parameter value is considered, the boundaries of the prior separated by the $\div$ symbol are reported. The $x$ and $y$ coordinates are expressed with respect to the position of the BCG-N. As in \citetalias{Bergamini_2021}, a Gaussian prior is assumed on the normalization value ($\sigma_{LT}^{ref}$) of the first scaling relation in \Eq\ref{eq.: Scaling_relations}. The mean and the standard deviation values of the Gaussian prior are quoted in square brackets. The total number of galaxies ($N_{gal}$) optimized through the scaling relations, and the reference magnitude value (${m_{\mathrm{F160W}}^{ref}}$) are also reported. 
    {\it Bottom:} For each free parameter of the reference lens model, we quote the median value and the 16-th and 84-th percentiles of the marginalized posterior distribution. 
    }    

	\label{table:inout_lensing}

\end{table*}

\begin{table}[h!]  
	\tiny
	\def\arraystretch{1.6}
	\centering    
	\begin{tabular}{|c|c|c|c|>{\centering\arraybackslash}m{9cm}|}
	   \hline
	   \multicolumn{4}{|c|}{\textbf{\normalsize Comparison between published lens models}}\\[2pt] 
	   \hline
	   \textbf{\normalsize Model} & \boldmath{\normalsize $N_\mathrm{images}$} & \boldmath{\normalsize $N_\mathrm{sources}$} & \boldmath{\normalsize $\Delta_{rms}\,\mathrm{[\arcsec]}$} \\[2pt] 
	   \hline
	   \textbf{This work} & \textbf{237} & \textbf{88} & \textbf{0.43} 
	   \cr
	   \hline
	   \citetalias{Bergamini_2021} & 182 & 66 & 0.40
	   \cr
	   \hline
	   \citetalias{Richard_2021} & 198 & 71 & 0.58
	   \cr
	   \hline
	   CATS & 116 & 41 & 0.67
	   \cr
	   \hline
	\end{tabular}
	\smallskip
    \caption{Comparison between our new lens model for \CL\ and other published models for the same cluster. {\normalsize $N_\mathrm{images}$} is the number of multiple images used as model constraints, {\normalsize $N_\mathrm{sources}$} is the number of background sources, and {\normalsize $\Delta_{rms}$} is the total root-mean-square displacement value between the observed and model-predicted image positions (see \Eq\ref{eq.: rms_lt}).
    }
	\label{table:lens_models} 
\end{table}

\begin{figure}
	\centering
	\includegraphics[width=1\linewidth]{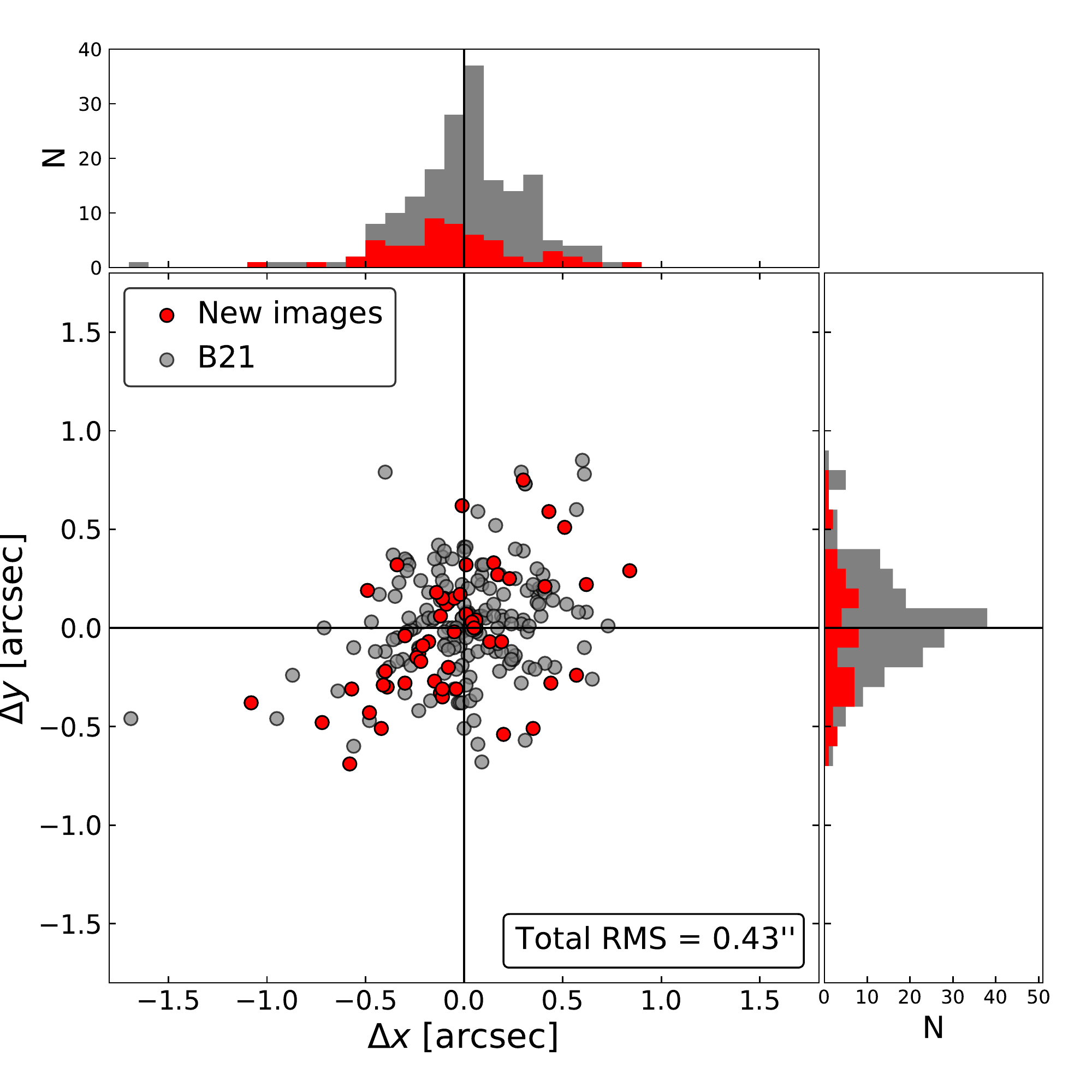}
	\caption{Displacements $\boldsymbol{\Delta_i}$ (see \Eq\ref{eq.: rms_lt}) along the $x$ and $y$ directions of the 237 observed multiple images used to optimize the reference lens model described in this work. Histograms show the displacement distribution along each direction. Gray circles correspond to the images in common with \citetalias{Bergamini_2021}, while the new images are plotted in red.}
	\label{fig:rms}
\end{figure}

\begin{figure*}
	\centering
	\includegraphics[width=0.925\linewidth]{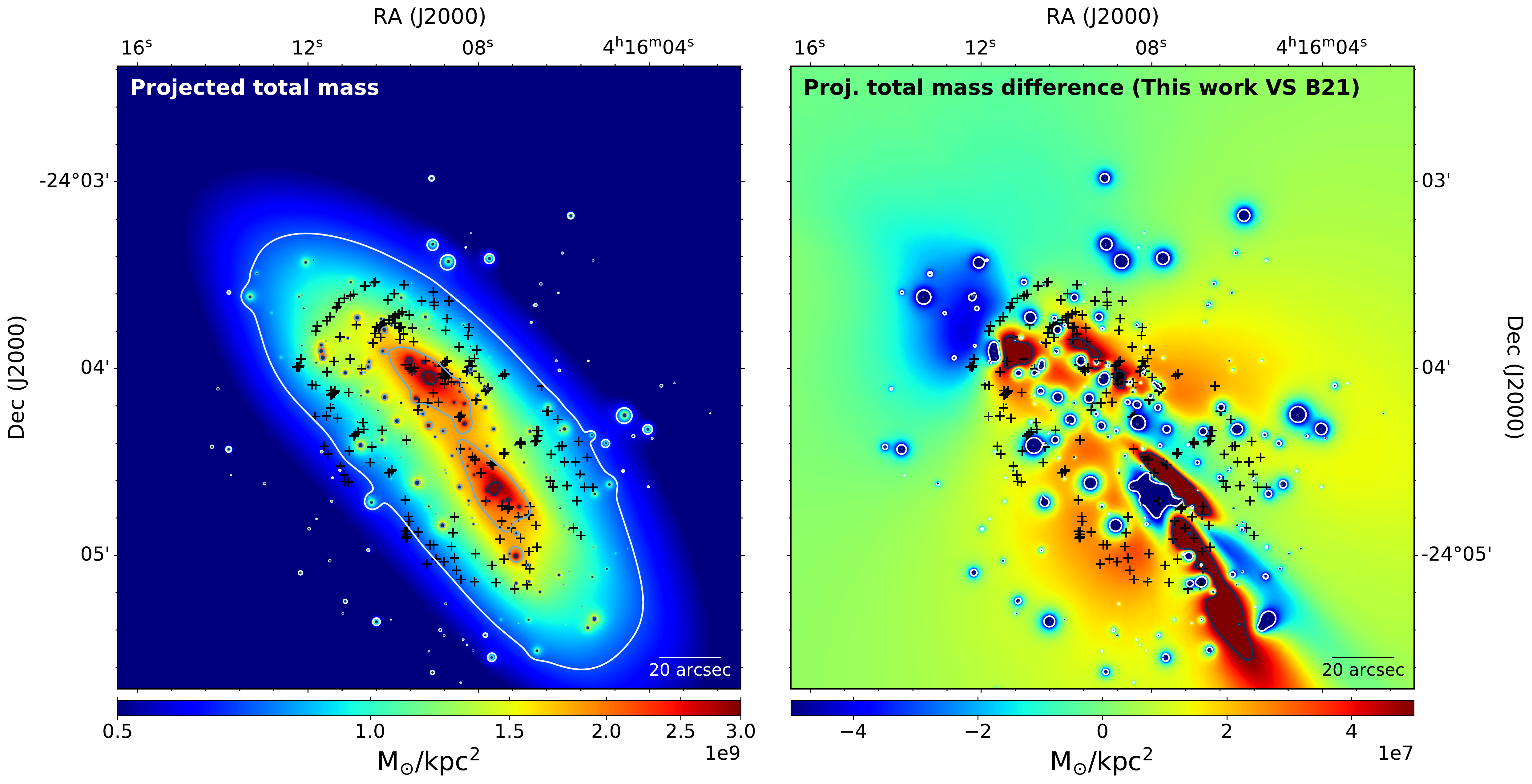}
	\caption{Projected total mass density distribution of \CL\ from the best-fit lens model presented in this work. {\it Left:} Projected total mass density distribution of \CL. Contour levels correspond to values of $ \left[0.70, 1.85, 3.00\right]\times 10^{9}\, \mathrm{M_{\odot}\,kpc^{-2}}$. {\it Right:} Difference between the cluster total mass density distributions inferred from the present reference lens model and that by \citetalias{Bergamini_2021}. This map is obtained by considering 500 realizations of the lens models randomly extracting samples of free parameter values from the MCMC chains. The procedure adopted to generate this map is detailed in \Sec\ref{sec:results}. Contour levels correspond to values of $\left[-5, 5\right]\times 10^{7}\, \mathrm{M_{\odot}\, kpc^{-2}}$ (i.e., the limits of the colorbar). The observed positions of the 237 multiple images used as model constraints are marked with black crosses.}
	\label{fig:mass}
\end{figure*}

\begin{figure*}
	\centering
	\includegraphics[width=0.925\linewidth]{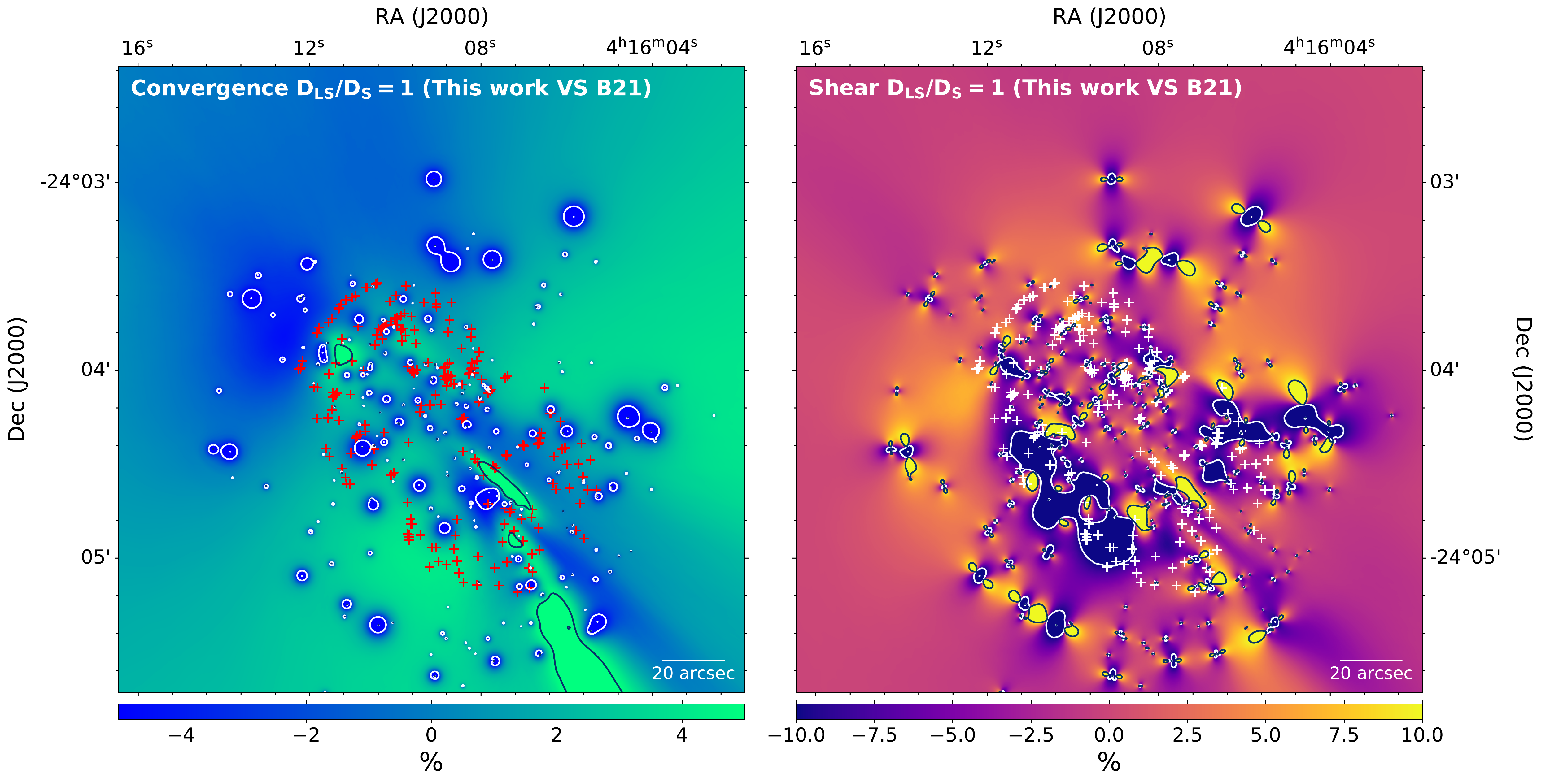}
	\caption{Percentage difference for the convergence (on the left) and the shear (on the right) distributions derived from our new reference lens model and that by \citetalias{Bergamini_2021}. To generate these maps, we use 500 random realizations of the lens models by extracting samples of free parameter values from the MCMC chains. For a detailed description on how the maps are created, see \Sec\ref{sec:results}. Contour levels correspond to values of $-$5\%, 5\% and $-$10\%, 10\%, respectively. These are the limits of the plot colorbars. The observed positions of the 237 multiple images used to constrain the lens model are marked with red/white crosses.}
	\label{fig:k_gamma}
\end{figure*}

\subsection{Multiple images}
\label{sec:multiple_images}
The strong lensing model of the galaxy cluster \CL\ developed in this work considers as constraints the point-like positions of the 182 multiple images used by \citetalias{Bergamini_2021}. In addition to those, 55 newly identified images are included in the catalog (see \Fig\ref{fig:hist} and \T\ref{tab:multiple_images}) and are briefly discussed below. 

Of the new images with respect to \citetalias{Bergamini_2021}, 50 are identified after cross-matching and complementing the spectroscopic catalog by \citetalias{Richard_2021}, considering only the images with the highest ($>1$) confidence quality flags (QF), with the samples from \cite{Vanzella_2020} and \citet{Mestric2022}. Going beyond the analysis presented in \citetalias{Richard_2021}, we identify multiply-lensed, nearly point-like substructures inside most of these images that are used as constraints in our model. The inclusion of these sub-knots in the lens model has proven to be extremely useful to reconstruct the fine details of the cluster gravitational potential and the positions of the critical lines \citep[see also][]{Grillo2016, Bergamini2022}. 
For all the new images, we then revisit the redshift value provided in \citetalias{Richard_2021} by cross-correlating and/or reanalyzing the MUSE spectra, extracted within customized apertures following the shape of the distorted images. In particular, we correct the redshift value of the system 202 (system 81 in the \citetalias{Richard_2021} catalog), that was incorrectly quoted equal to 1.827 by \citetalias{Richard_2021}, instead of 2.091 (see \T\ref{tab:multiple_images}).

Two of the remaining five images, with ID 211b and 211c, were identified by \cite{Vanzella_2020b}.
In addition, we securely associate a third counter image (205c) with the image family 205 (family 91 in \citetalias{Richard_2021}), that was not considered by \citetalias{Richard_2021}. 
The last two new images belong to Sys16, a galaxy-galaxy strong lensing system around the cluster galaxy identified as Gal-8785 (see \Fig\ref{fig:RGB} and \Sec\ref{sec:results}). A careful inspection of the multiple images around this galaxy and of the lens model predictions have allowed us to correct a few inaccurate associations (with no impact on the overall cluster lens model) assumed in \citetalias{Bergamini_2021}. As a result, two additional images of Sys16 are included near the galaxy Gal-8785 (see \T\ref{tab:multiple_images} and the bottom-left panel of \Fig\ref{fig:FM_sys16}).

Compared to the \citetalias{Richard_2021} multiple image lensing catalog, this work includes 75 new images. On the other hand, 36 images from \citetalias{Richard_2021} are not considered here based on low confidence redshift measurements (QF$<$1) and/or unclear identifications based on the HST imaging.

\subsection{Mass parameterization}

\citetalias{Bergamini_2021} explored several total mass parameterizations by varying the number of large-scale DM halos and testing the impact of additional external shear and convergence terms. In this work, we adopt for \CL\ the total mass parameterization of the reference model in \citetalias{Bergamini_2021}, which was significantly favored by all the considered statistical estimators. We refer the reader to that reference for a detailed overview, and provide here a brief summary. 
\LT\ implements a parametric approach to model the total mass distribution of galaxy clusters. This means that the total gravitational potential of \CL\ is divided into the following sum of different components:

\begin{equation}
    \label{eq.: pot_dec}
    \phi_{tot}= \sum_{i=1}^{N_h}\phi_i^{halo}+\sum_{j=1}^{N_{gas}}\phi_j^{gas}+\sum_{k=1}^{N_g}\phi_k^{gal}+\phi_{foreg}.
\end{equation}

\noindent The first sum runs over the mass density profiles used to parameterize the cluster-scale halos of the cluster (mainly made of dark matter), the second one describes the contribution of the hot gas to the total cluster mass, the third one takes into account the total mass distribution of the cluster member galaxies (the sub-halo component), and finally, the last term is used to represent a foreground galaxy, at $z=0.112$, residing in the southern region of \CL\ (see \Fig\ref{fig:RGB}). 
In the model we are presenting, each mass component ($\phi_i^{halo}$, $\phi_j^{gas}$, $\phi_k^{gal}$, and $\phi_{foreg}$) is described by a dual pseudo-isothermal elliptical mass profile \citep[dPIE,][]{Limousin_lenstool, Eliasdottir_lenstool, Bergamini_2019} that is characterized by seven free parameters: two parameters define the position on the sky ($x$, $y$), two correspond to the ellipticity ($e=\frac{a^2-b^2}{a^2+b^2}$, where $a$ and $b$ are the values of the semi-major and semi-minor axes of the ellipsoid, respectively) and the position angle ($\theta$, computed counterclockwise from the west direction), while the last three parameters, $\sigma_{0}$, $r_{core}$, and $r_{cut}$, are the central velocity dispersion, the core radius, and the truncation radius, respectively. In passing, we note that, instead of using $\sigma_{0}$, \LT\ implements a scaled version of this quantity, identified as $\sigma_{LT}$, such that $\sigma_{LT}=\sigma_{0} \sqrt{2/3}$.

The cluster-scale component ($\phi_i^{halo}$) of our new lens model is parametrized by three non-truncated elliptical dPIE profiles. Two are centered on the brightest cluster galaxies, BCG-N and BCG-S in \Fig\ref{fig:RGB}, while the position of the third one is free to vary in the southern part of the cluster, and it is necessary to provide second-order corrections to the total cluster mass distribution around the southern BCG. An extra circular and non-truncated dPIE profile is included in the lens model to account for a small over-density of galaxies in the North-East region of the cluster (around Sys4a,b in \Fig\ref{fig:RGB}). 

For the values of the dPIE parameters describing the hot-gas component of \CL\ ($\phi_i^{gas}$), we make use of the results by \cite{Bonamigo_2017, Bonamigo_2018}. In particular, \cite{Bonamigo_2018} found that the total hot-gas content of \CL, inferred from the Chandra X-ray observations, can be well characterized using four elliptical dPIE profiles. Since the values of the parameters of these profiles are kept fixed, the hot-gas component does not introduce any extra free parameter in the lens model. 

The cluster member galaxies ($\phi_j^{gal}$) are modeled with singular, circular dPIE profiles, for which the velocity dispersion, $\sigma^{gal}_{LT,i}$, and truncation radius, $r^{gal}_{cut,i}$, values scale with that of the galaxy luminosity, $L_i$, according to the following relations (which are used to sensibly reduce the number of free parameters of the lens model):

\begin{equation}
    \sigma^{gal}_{LT,i}= \sigma^{ref}_{LT} \left(  \frac{L_i}{L_{ref}} \right)^{\alpha},\quad
    r^{gal}_{cut,i}= r^{ref}_{cut} \left(  \frac{L_i}{L_{ref}} \right)^{\beta_{cut}}.
    \label{eq.: Scaling_relations}
\end{equation}
\medskip

\noindent In these equations, the reference luminosity, $L_{ref}$, corresponds to the BCG-N magnitude in the \Hubble\ F160W band ($m_{\mathrm{F160W}}^{ref}=17.02$). Following \citetalias{Bergamini_2021}, we fix $\alpha=0.3$ and $\beta_{cut}=0.6$, while a Gaussian prior with a mean value of 248\,km s$^{-1}$ and a standard deviation value of 28\,km s$^{-1}$ is assumed on the reference velocity dispersion value, $\sigma^{ref}_{LT}$. All these values are inferred from the measured inner stellar kinematics of 64 cluster member galaxies, obtained by exploiting the high-quality MUSE data (see \citetalias{Bergamini_2021}). An uniform prior between 1\arcsec\ and 50\arcsec\ is assumed on the $r^{ref}_{cut}$ value. 
As in \citetalias{Bergamini_2021}, we model the galaxy identified as Gal-8971 with a singular dPIE profile, outside of the cluster member scaling relations (thus resulting in four additional free parameters). The total mass distribution of this galaxy is mostly responsible for the formation of a galaxy-scale strong lensing event composed of four multiple images (separated by $\sim 1\arcsec$) of the same background source at $z=3.221$ \citep{Vanzella_ID14}. 

Finally, the total mass contribution of the bright foreground galaxy at $z=0.112$, $\phi_{foreg}$, is modeled as a first approximation at the cluster redshift \citep[as similarly done in previous works, see e.g.,][]{Richard_2014, Grillo_2015, Kawamata_2016, Caminha_macs0416}. 
Located in the southern region of the cluster (see \Fig\ref{fig:RGB}), the galaxy lies angularly very close to the families 36 and 37, impacting the observed positions of the multiple images belonging to these families. This foreground perturber is parameterized using a singular, circular dPIE profile with its position fixed on the centroid of the galaxy light emission. This extra profile adds two more free parameters to the lens model.

The lens model includes a total of 30 free parameters and 298 constraints, which corresponds to 268 dof. On the top of \T\ref{table:inout_lensing}, we summarize the fixed values and the priors assumed for the parameter values of the mass profiles included in the lens model. On the bottom of the same table, we show instead the median values and confidence intervals obtained from the sampling. 

\section{Results}
\label{sec:results}
In \Fig\ref{fig:rms}, we show the displacements, $\boldsymbol{\Delta}_i$, along the $x$ and $y$ directions, between the observed and model-predicted positions of the multiple images used to constrain the lens model. Of the 237 images, only four have a $\left\|\boldsymbol{\Delta}_i\right\|$ value larger than 1\arcsec. One of them, identified as 101c in the image catalog (with $z=4.2994$ and $\left\|\boldsymbol{\Delta}_{101c}\right\|=1.75\arcsec$), resides between two background galaxies, the first, at a projected distance of $\sim2.5$\arcsec\, has $z=0.5377$ and a total magnitude of $m_\mathrm{F160W}=20.76$, while the second one, at a projected distance of $\sim0.8$\arcsec, has $z=0.5660$ and $m_\mathrm{F160W}=24.46$. None of these two galaxies is included in the lens model, and this can marginally affect the $\Delta_{rms}$ value of the images in their vicinity. 
A second image, labeled with 102c, with $\left\|\boldsymbol{\Delta}_{102c}\right\|=1.05\arcsec$ and $z=6.0644$, lies in the same region of the cluster just 3.8\arcsec\ away from 101c. Another image, identified as 1a, with $z=3.238$ and $\left\|\boldsymbol{\Delta}_{1a}\right\|=1.04\arcsec$, is located at a projected distance of 6.2\arcsec\ from a background spiral galaxy at $z=0.5277$ not included in the lens model.
We note that considering in the modeling background galaxies is not straightforward, since their position and magnitude values are affected by the lensing effect of the cluster. The inclusion of these galaxies, effectively at the redshift of the cluster, could introduce potential biases. For instance, the real (i.e. delensed) positions of the background galaxies, at their correct redshifts, depend on the optimized total mass distribution of the cluster. We also note that the work of \cite{Chirivi_2018} showed that the perturbing lensing effect of foreground galaxies is more important than that of background galaxies. For all these reasons, the background galaxies mentioned above have been for now excluded from the modeling.
Finally, the fourth multiple image, denominated 210.4b ($\left\|\boldsymbol{\Delta}_{210.4b}\right\|=1.15\arcsec$), consists in a Ly$\alpha$ emitter at $z=6.149$. This image forms close (just 1.3\arcsec \,away) to a bright cluster galaxy, identified as Gal-7955, with $m_\mathrm{F160W}=21.29$. Since the total mass of this cluster member is approximated through the lens model scaling relations (see \Eq\ref{eq.: Scaling_relations}), a deviation of its real mass distribution from the predictions of the best-fit relations could justify the displacement of this image. 

\begin{figure}[t]
	\centering
	\includegraphics[width=1\linewidth]{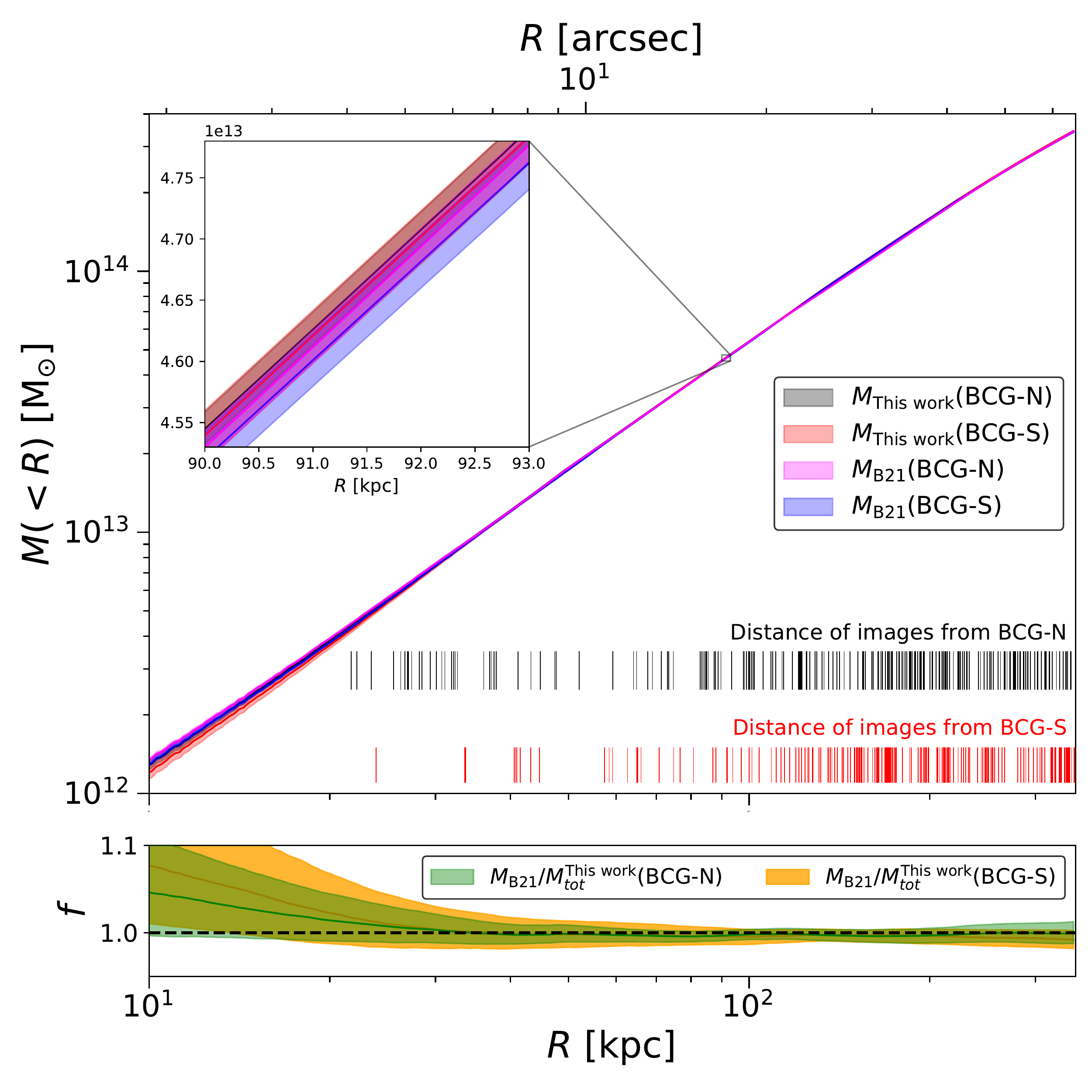}
	\caption{Comparison of the cumulative projected total mass profiles of \CL\ from this work and that by \citetalias{Bergamini_2021}. {\it Top:} Cumulative projected total mass profiles of \CL \, as a function of the projected distance $R$ from the northern and southern BCG, obtained from our new reference lens model ($M_\mathrm{This\ work}$(BCG-N), $M_\mathrm{This\ work}$(BCG-S)) and from the model by \citetalias{Bergamini_2021} ($M_\mathrm{B21}$(BCG-N), $M_\mathrm{B21}$(BCG-S)). The distances of the observed multiple images from the BCGs, used in this work, are plotted using small vertical bars. {\it Bottom:} Ratio between the same mass profiles as derived from our new reference model and that by \citetalias{Bergamini_2021}.}
	\label{fig:mass_profile}
\end{figure}

\begin{figure}[t]
	\centering
	\includegraphics[width=1\linewidth]{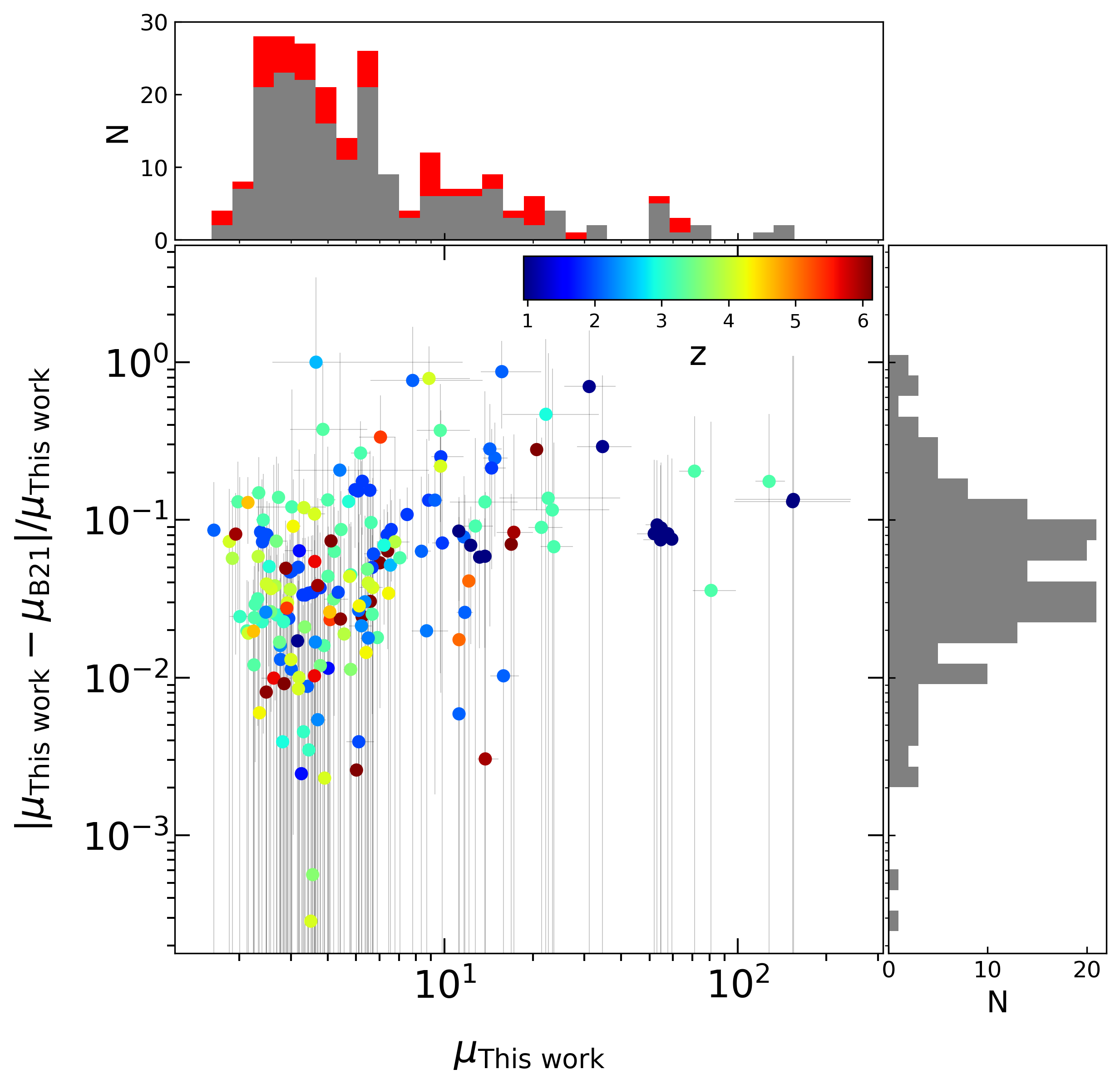}
	\caption{Absolute relative difference between the absolute magnification factors obtained from the current and the \citetalias{Bergamini_2021} lens models as a function of the magnification values from this work. The data points represent the different observed multiple images included both in the \citetalias{Bergamini_2021} and the current lens models. The data points are color-coded according to the redshift of the background sources. The absolute magnification value, $\mu$, corresponds to the median absolute magnification computed considering 100 different realizations randomly extracted from the final MCMC chains. The errors in the magnification values are computed from the 16th and 84th percentiles. The distributions of the absolute image magnifications obtained from the current model and their absolute relative differences with the values obtained by \citetalias{Bergamini_2021} are shown on the top and right histograms, respectively. The red histogram on the top shows the absolute magnification distribution of the newly identified images (i.e., not used by \citetalias{Bergamini_2021}).}
	\label{fig:ampli_dist}
\end{figure}

\begin{figure*}
	\centering
	\includegraphics[width=1\linewidth]{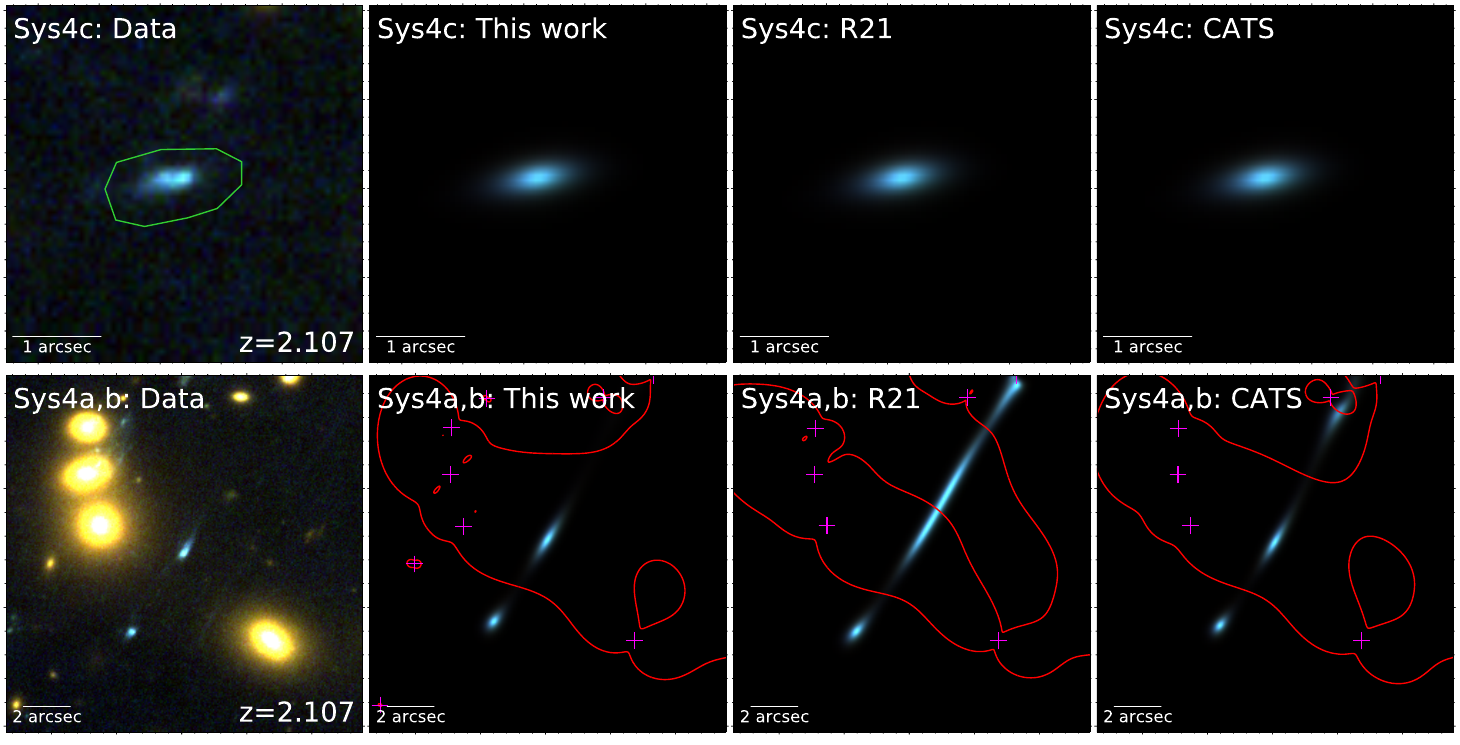}
	\caption{Forward modeling reconstruction of the multiple image system 4, highlighted in cyan in \Fig\ref{fig:RGB}. The RGB images in the leftmost column are obtained by combining the \Hubble\ F435W, F606W, and F814W filters. The next three columns, from left to right, show the model predicted images obtained by using our forward modeling code {\scshape GravityFM} (see \Sec\ref{sec:results}) and adopting the deflection maps of our new reference model, the model by \citetalias{Richard_2021}, and the CATS model, respectively. The green polygon in the top-left image contains the pixels associated to the single image (4c) that is exploited by {\scshape GravityFM} to determine the best-fit parameter values of the background source, while the other images (4a and 4b) are used as test images and predicted a-posteriori. The cluster tangential critical lines, computed at the source redshift, are plotted in red. Magenta crosses mark the positions of the cluster member galaxies included in each model.}
	\label{fig:FM_sys4}
\end{figure*}

\begin{figure*}
	\centering
	\includegraphics[width=1\linewidth]{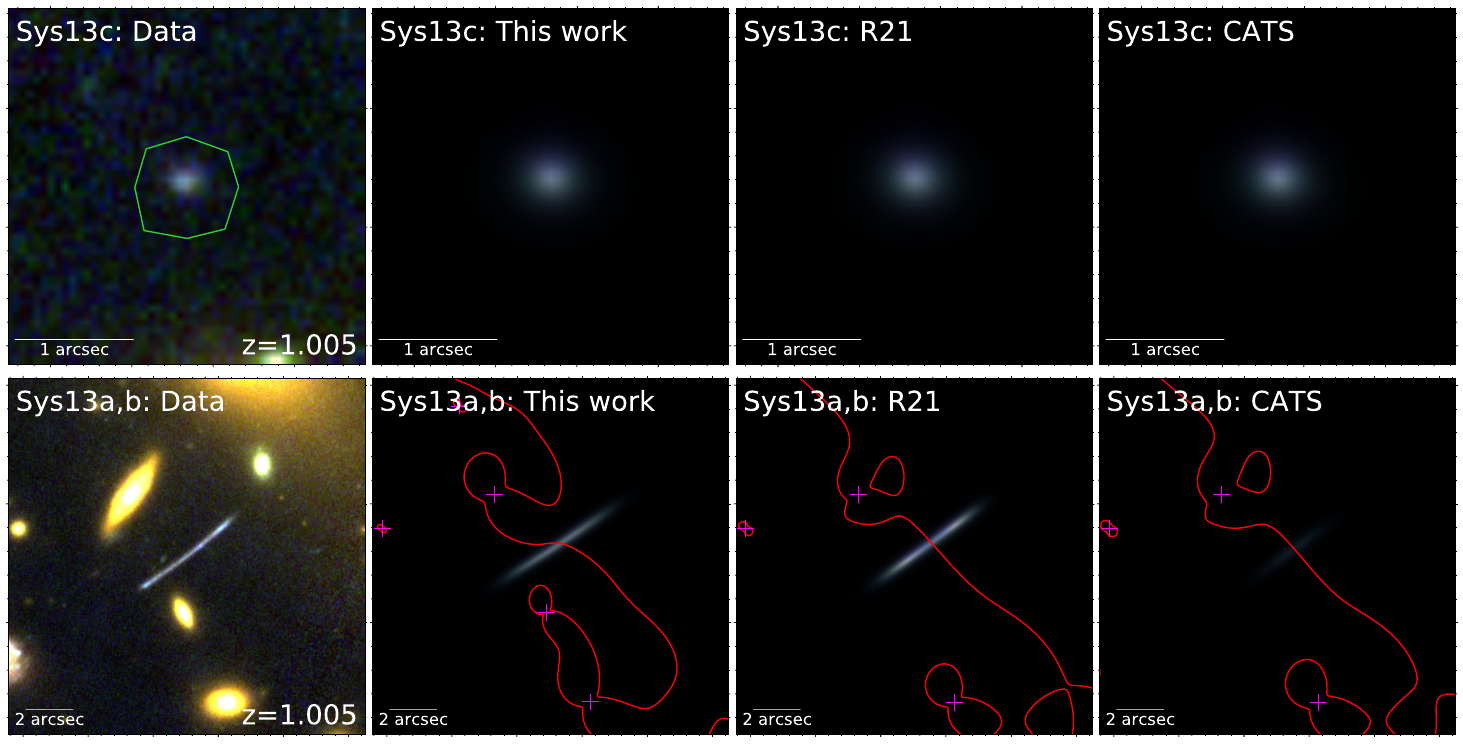}
	\caption{Same as \Fig\ref{fig:FM_sys4} for Sys13, highlighted in yellow in \Fig\ref{fig:RGB}.}
	\label{fig:FM_sys13}
\end{figure*}

\begin{figure*}
	\centering
	\includegraphics[width=1\linewidth]{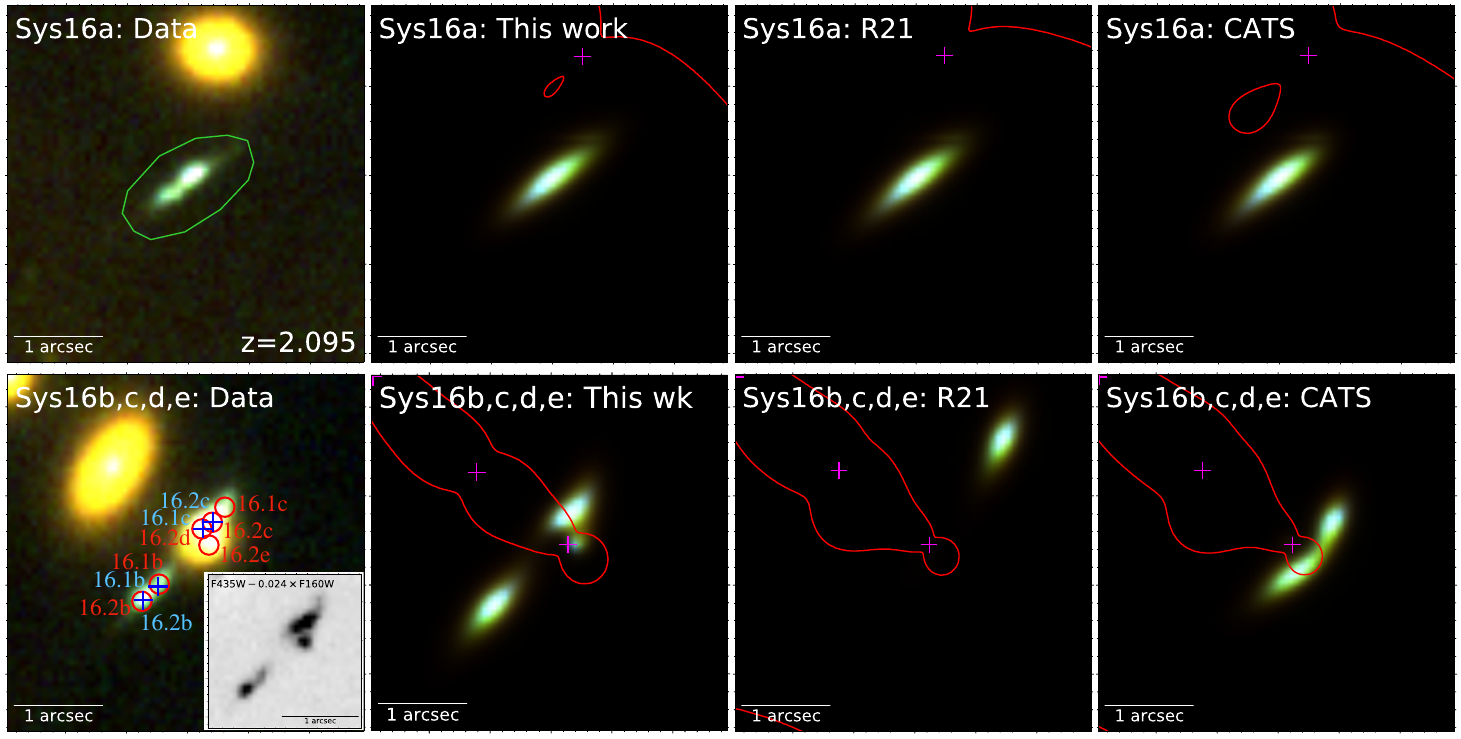}
	\caption{Same as \Fig\ref{fig:FM_sys4} for Sys16, highlighted in orange in \Fig\ref{fig:RGB}. In the bottom-left panel, we mark with blue crosses the observed positions of the multiple images adopted by \citetalias{Bergamini_2021} and with red circles the image configuration included in the current model (see \Sec\ref{sec:multiple_images}). The small inset on the bottom left panel shows a liner combination of two \Hubble\ filters to subtract the cluster member light contribution.}
	\label{fig:FM_sys16}
\end{figure*}

\begin{figure*}
	\centering
	\includegraphics[width=1\linewidth]{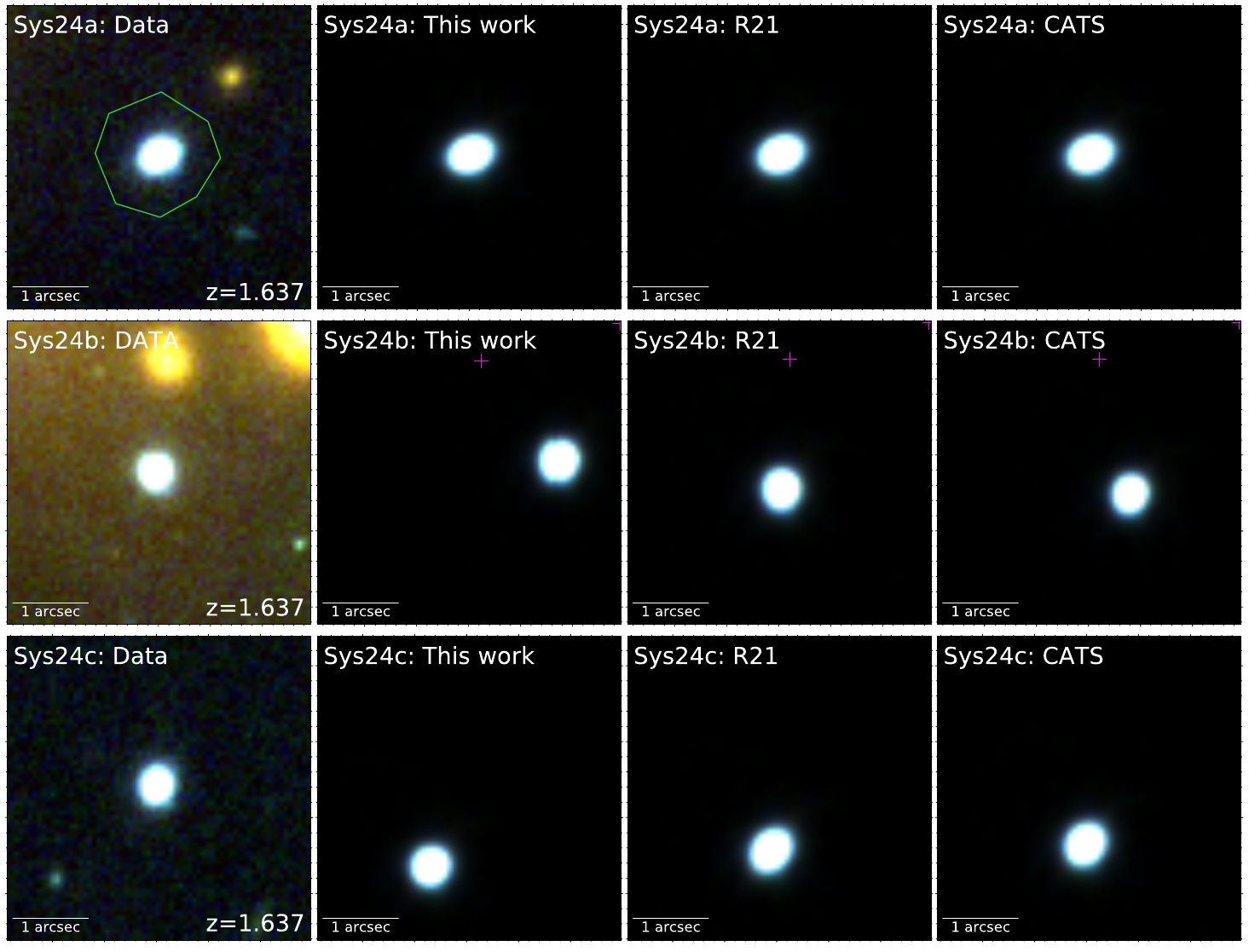}
	\caption{Same as \Fig\ref{fig:FM_sys4} for Sys24, highlighted in pink in \Fig\ref{fig:RGB}.}
	\label{fig:FM_sys24}
\end{figure*}

\begin{figure*}
	\centering
	\includegraphics[width=1\linewidth]{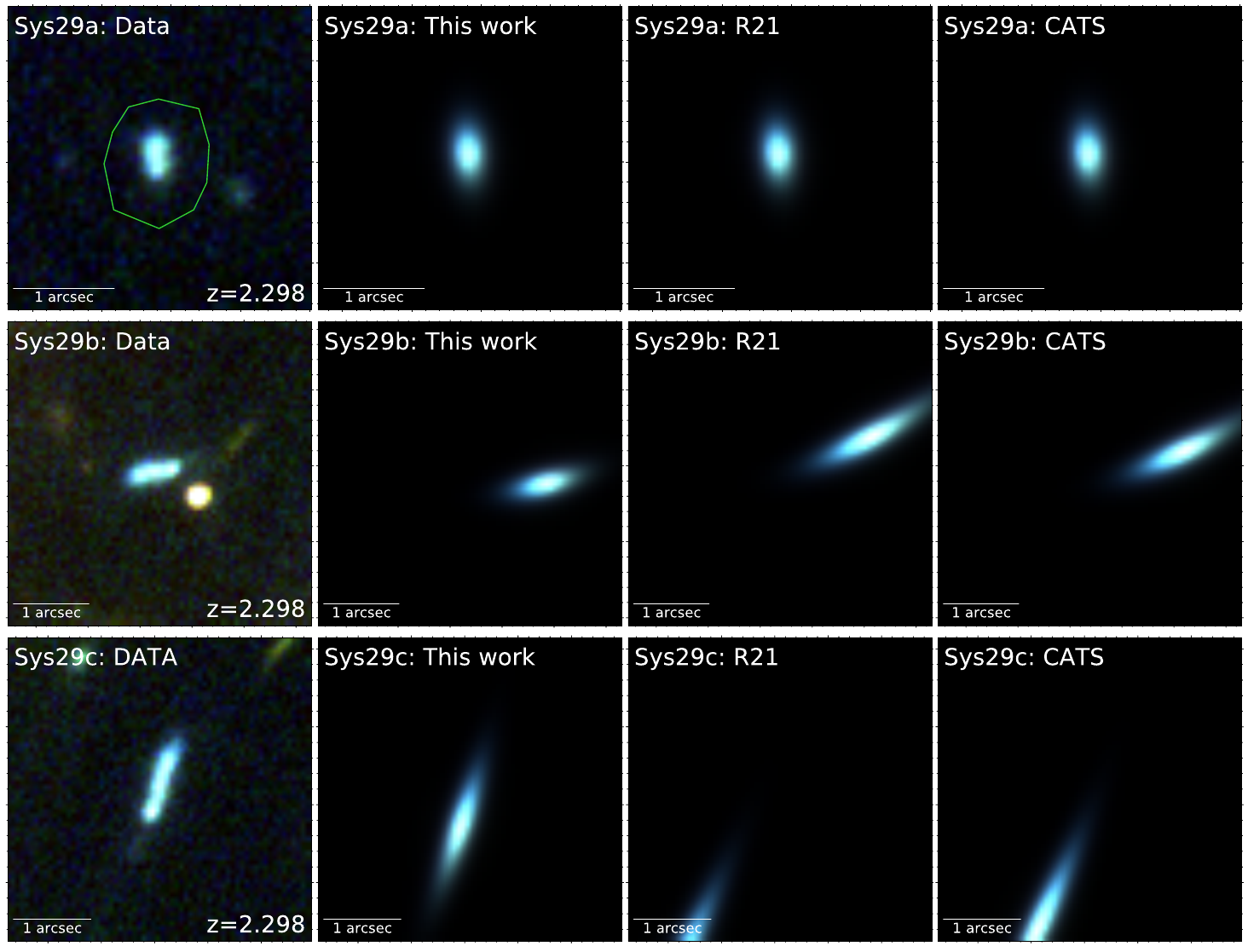}
	\caption{Same as \Fig\ref{fig:FM_sys4} for Sys29, highlighted in green in \Fig\ref{fig:RGB}.}
	\label{fig:FM_sys29}
\end{figure*}

The global $\Delta_{rms}$ value of the lens model stands at 0.43\arcsec; about 80\% of the multiple images have $\left\|\boldsymbol{\Delta}_i\right\|<0.5\arcsec$, while more than 68\% of them have $\left\|\boldsymbol{\Delta}_i\right\|<\Delta_{rms}=0.43\arcsec$.

The model we are presenting is characterized by a sample of secure multiple images that is $\sim\!30\%$ larger than the already large one in \citetalias{Bergamini_2021} (see \Fig\ref{fig:image_distribution}). The additional multiple images allow us to better constrain the fine details of the \CL\ total mass distribution. On the right panel of \Fig\ref{fig:mass}, we compare the total projected mass distribution obtained using the current model with that presented by \citetalias{Bergamini_2021}. To account for the statistical errors, we consider 500 realizations of both models randomly extracting samples of free parameter values from the \LT\ MCMC chains. The total mass map associated with each random realization of the \citetalias{Bergamini_2021} model is then subtracted to one of those obtained using our new model, in order to get distributions with 500 values of total mass difference in every pixel. In the right panel of \Fig\ref{fig:mass}, we plot a map showing for each pixel the median values of these distributions. A similar procedure is also applied to obtain the two panels of \Fig\ref{fig:k_gamma}, where we compare the convergence and shear maps, computed fixing the ratio of the lens-source to observer-source distances equal to 1. We note that differently from \Fig\ref{fig:mass}, we consider here distributions of normalized values. These are obtained by dividing each of the 500 convergence/shear differences by the median map of the 500 realizations of our new model.

Both the right panel of \Fig\ref{fig:mass} and the left panel of \Fig\ref{fig:k_gamma} illustrate that in the core of \CL, where we observe the multiple images that constrain the lens models, the \citetalias{Bergamini_2021} and the new models are characterized by very similar total mass distributions. In that region, the mass difference between the two models is mostly smaller than 5\%, corresponding to just a few tens of millions of solar masses per square kpc for the projected total mass density. As a reference, we show on the left panel of \Fig\ref{fig:mass} the total projected mass distribution of \CL\ obtained from our new best-fit lens model. This small difference is also reflected in the small offsets visible in \Fig\ref{fig:mass_profile}, which shows the cumulative total mass distributions of the cluster, from the model by \citetalias{Bergamini_2021} and that we are presenting here, as a function of the distance from the two BCGs. This plot highlights differences of about 5\% close to BCG centers (there are no multiple images constraining the lens mass in those regions), decreasing to less than 1\% at distances larger than 20\,kpc, where the first multiple images are observed. As shown in the right panel of \Fig\ref{fig:k_gamma}, slightly larger differences, but mostly below 10\%, are found between the shear maps of the new and \citetalias{Bergamini_2021} lens models. When comparing with the \citetalias{Richard_2021} and CATS models (more details about these models are provided below), we find that the resulting cumulative total profiles are consistent in the regions where multiple images are identified (between $\sim$20 and $\sim$200 kpc from the BCG-N and the BCG-S). The differences outside of these regions are about 10\%.

In \Fig\ref{fig:ampli_dist}, we show the absolute relative difference between the absolute magnification values obtained by using our current model and that by \citetalias{Bergamini_2021}. This analysis demonstrates that both models predict very similar magnification values for most of the multiple images. In fact, of the 182 multiple images shared with the model by \citetalias{Bergamini_2021}, 95 (141) have absolute relative differences in the magnification values smaller than $0.05$ ($0.1$). As expected, larger differences are observed for the most magnified images. In fact, $\sim 60\%$ of the images with a magnification value lower than 10 have an absolute relative difference lower than $0.05$, while less than $20\%$ of the images with a magnification larger than 10 have an absolute relative difference lower than $0.05$. These most magnified images lie closer to the cluster critical lines and are thus more affected by the small differences between the two lens models.

Even though the $\Delta_{rms}$ value, defined in \Eq\ref{eq.: rms_lt}, is a valid figure of merit to quantify the overall goodness of a lens model, this estimator is computed considering only the point-like positions of the observed and model-predicted multiple images. As a result, lens models with comparable $\Delta_{rms}$ values might not necessarily be equally good at reproducing the distorted surface brightness distribution of multiply lensed sources. To make some progress on this, we present a comparison between our new model and two previously published lens models for \CL, with the aim of testing their ability in predicting the details of the extended emission of five systems of multiple images. 
The first public model was presented by \citetalias{Richard_2021}. It counts 198 spectroscopically confirmed multiple images from 71 sources, and it has $\Delta_{rms}=0.58$\arcsec. Based on the publicly available files of the \citetalias{Richard_2021} lens model, \CL\ is parameterized with 3 cluster scale dark matter halos, 98 cluster members, 97 of which are modeled within the scaling relations, and 2 additional (one in the foreground and one in the background) galaxies modeled separately.
The second available model was developed by the CATS \citep[Clusters As TelescopeS,][]{Jauzac_2014, Richard_2014, Jauzac_2015} team, which is known as the v4 CATS lens model (labeled here as CATS model). The latter is constrained by 116 multiple images from 41 sources, and it provides $\Delta_{rms}=0.67$\arcsec. The CATS lens model also includes 3 cluster scale dark matter halos and 98 member galaxies (2 of which are modeled outside of the scaling relations) and a foreground galaxy modeled separately. We note that neither of these lens models include any information about the gas content of the cluster, based on X-ray observations, and the stellar kinematics of the cluster galaxies to accurately constrain the sub-halo component.
In \T\ref{table:lens_models}, we summarize the main characteristics of the compared lens models, while in \Fig\ref{fig:image_distribution} we plot the cumulative distributions of the number of multiple images, used as constraints for the different models, as a function of their distance from the northern BCG.

The five analyzed systems of extended multiple images are marked with colored rectangles in \Fig\ref{fig:RGB} and are selected according to the following criteria:

\begin{itemize}
    \item They are included in the catalogs of multiple images used as constraints by all lens models we are comparing.
    \item The morphology of the observed extended images is sufficiently simple to be well approximated by a single Sérsic light model on the source plane.
    \item They are located in different regions of the cluster, and they are distributed across the whole cluster field-of-view: one in the North, two at the center, and two in the South. 
\end{itemize}

The reconstruction of the surface brightness distribution of the multiple images is entrusted to a novel forward modeling code we have developed, named {\scshape GravityFM}. {\scshape GravityFM} is a python-based software that, through a high-level interface, allows for the optimization of the parameter values of light models chosen to represent background sources from the observed, distorted, and magnified surface brightness distributions in a given filter of their multiple images. In particular, {\scshape GravityFM} implements a Bayesian approach to minimize the residuals between the observed and model-predicted surface brightness of the multiple images. For a detailed description of this software, we refer to the reference paper by \ci{Bergamini et al. (in preparation)}. 
By adopting the deflection maps of the best-fit lens models we are comparing, we use {\scshape GravityFM} to find the best-fit parameter values of the Sérsic profile, that we use to describe the background source, corresponding to one of the observed multiple images of a given system. In all the cases, we assume that the background source can be well approximated by a single Sérsic model with an index value $n=1$, while the following parameter values are left free to vary: the coordinates, $X$ and $Y$, of the center on the source plane, the effective radius $R_e$, the axis ratio $q$, the position angle $\theta$, and the total emitted flux $F$. The image pixels considered in the optimization process are encircled in green in Figs. \ref{fig:FM_sys4}, \ref{fig:FM_sys13}, \ref{fig:FM_sys16}, \ref{fig:FM_sys24}, and \ref{fig:FM_sys29}. For each system, these images are the least distorted ones and without significant contamination from other bright sources in their vicinity. 
First, the selected single image of a system is used to optimize the parameter values of the corresponding background source, then we use that reconstructed source and {\scshape GravityFM}, with the deflection maps obtained from each different lens model, to predict a posteriori the surface brightness distributions of the other counter-images.
To create the color multiple images displayed in the figures, we perform for each system three forward modeling optimizations using the F814W, F606W, and F435W \Hubble\ filters. The model predictions in each band are then combined to obtain the RGB images.

As expected, the \citetalias{Richard_2021}, the CATS and our new lens models are equally good at reproducing the chosen first images of every system. This is not surprising since these images are located in regions of low magnification for all lens models and, by changing the values of the free parameters of the Sérsic profile of the background sources, satisfactory fits to the data are achieved. We note that the different lens models reconstruct sources with different best-fit values for the parameters (i.e., centroid, ellipticity, and intensity) of their surface brightness distribution. It is not straightforward instead if the optimized source can also reproduce well, a-posteriori, the position, shape and flux of the other extended images of each system, because their observations are not considered during the reconstruction of the corresponding sources.

In \Fig\ref{fig:FM_sys4}, we show the results for Sys4. In this case, our lens model is the only one that is able to reproduce the correct observed configuration of the two multiple images identified as Sys4a,b. On the contrary, the \citetalias{Richard_2021} and CATS models predict the formation of several additional bright images, some of them in the form of an extended arc, that are not observed in the \Hubble\ images. 

Sys13, illustrated in \Fig\ref{fig:FM_sys13}, is composed of three multiple images, two of which (a and b) merge into an extended arc that crosses the main cluster critical line at the source redshift ($z=1.005$). For this system, both our and the \citetalias{Richard_2021} models successfully predict the observed arc configuration, while in the CATS model the arc is so faint that it is just barely visible. 

In Sys16 (\Fig\ref{fig:FM_sys16}), the test extended images (i.e., b, c, d, and e) form around a cluster galaxy with $m_\mathrm{F160W}=21.48$. We note that this galaxy is included in all the lens models and its total mass parameter values are optimized within the adopted cluster member scaling relations (e.g., see \Eq\ref{eq.: Scaling_relations}). Remarkably, only our lens model can accurately reproduce the observed number of multiple images and their detailed surface brightness distributions. 

Finally, while all the lens models perform fairly well in reproducing the images b and c of Sys24, as shown in \Fig\ref{fig:FM_sys24}, our model provides the best predictions for the positions and shapes (i.e., ellipticity and orientation) of the two test images b and c of Sys29, as visible in \Fig\ref{fig:FM_sys29}.

These results further demonstrate the importance of including the stellar kinematic information of the cluster member galaxies to accurately constrain the sub-halo scaling relations, following \citet{Bergamini_2019} and \citetalias{Bergamini_2021}.
Recent comparisons between lensing models of massive clusters, such as \CL, and state-of-the art hydrodynamical simulations have shown a significant discrepancy between the observed and simulated probability of clusters to produce galaxy-galaxy strong lensing (GGSL) events \citep[e.g.,][]{Meneghetti_2020, Meneghetti_2022, Ragagnin_2022}. This result can be interpreted as observed cluster members being more compact than their simulated analogs. Given its accuracy in reproducing the observed features of the multiple images lensed by single cluster members, this model will be used as a basis for further, deeper studies on GGSL systems (\ci{Granata et al. in preparation}), which may provide us with an independent probe of the compactness of the member galaxies.

\section{Conclusions}
\label{sec:conclusions}
We have presented our latest high-precision strong lensing model for the galaxy cluster \CL\ ($z=0.396$), based on the largest spectroscopic sample of multiple images available to date. We have taken advantage of panchromatic \Hubble\ observations in combination with deep, high-quality MUSE spectroscopic data to identify 237 multiple images, 55 more than in \citetalias{Bergamini_2021} (and 75 new images with respect to \citetalias{Richard_2021}), from 88 background sources, as well as a pure and complete sample of cluster member galaxies. The multiple images cover a redshift range between 0.94 and 6.63 and are uniformly distributed across all the cluster field of view. The model incorporates the contribution to the cluster total mass of the baryonic hot-gas component \citep[derived from the \textit{Chandra} X-ray data in][]{Bonamigo_2018} and includes the measured stellar velocity dispersions of 64 cluster galaxies (corresponding to approximately 30\% of the total number of selected cluster members) to accurately characterize the sub-halo total mass component of the cluster. This new dataset makes M0416 arguably the best known cluster lens to date. 

By comparing our new results with those from previously published models for \CL, we have found that our model outperforms the others in terms of ability to reconstruct the observed positions, shapes, and magnifications of the multiply imaged sources. Thus, other than to robustly characterize the total mass distribution of the cluster down to the galaxy scale, our model can provide accurate and precise magnification maps, that are crucial for the study of the intrinsic physical properties of faint, high-redshift sources magnified by the lensing cluster \citep{Mestric2022}. These low-luminosity sources, the progenitors of the galaxies in the local Universe, may play an important role in the cosmic re-ionization, making them interesting targets for the James Webb Space Telescope (JWST) telescope. 
Our main results can be summarized as follows:

\begin{enumerate}
    \item The total root-mean-square separation between the observed and model-predicted positions of the 237 multiple images is $\Delta_{rms}=0.43\arcsec$. This value is very similar to the previous value by \citetalias{Bergamini_2021}, despite a $\sim 30\%$ increase in the number of  multiple images used to constrain the model. Moreover, the distance between the observed and predicted positions, $\left\|\boldsymbol{\Delta}_i\right\|$, is lower than the $\Delta_{rms}$ value for $\sim 69\%$ of the images.
    \\
    \item The projected total mass density distribution, the cumulative total mass profile, and the convergence and the shear maps obtained from our new model are highly consistent with those obtained by \citetalias{Bergamini_2021}. This result, in combination with a nearly unchanged $\Delta_{rms}$ value and a significantly larger sample of multiple images, demonstrates that our modeling parameterization is well suited to robustly characterize the total mass distribution of \CL.
    \\
    \item Thanks to a novel forward modeling code developed in this work, we have tested that our new lens model is able to faithfully reproduce the extended and distorted surface brightness distributions of several observed multiple images. In this context, our new high-precision model outperforms the results of previously published lens models (based on smaller samples of multiple images, in some cases not spectroscopically confirmed, and different modeling prescriptions for the total mass of cluster members) at reproducing the observed positions, shapes, and magnifications of the extended lensed systems.
    
\end{enumerate}

The \CL\ lens model presented here will be made publicly available with the publication of this paper through our newly developed Strong Lensing Online Tool, a.k.a. \SLOT\ \citep[][]{Bergamini2022}. With a simple graphical interface, \SLOT\ offers the opportunity for non-lensing experts to access all the results and to exploit the predictive power of our models for their studies, specifically those based on upcoming JWST observations of M0416, which will reveal yet more multiple images. For example, \SLOT\ can quickly predict all the multiple images of a given source, computing also associated quantities (such as positions, magnification factors, time delays) with statistical errors obtained from the model MCMC chains. Moreover, it can be used to create maps of magnification, projected total mass, deflection angle, etc. These high-level products can, for instance, be used in combination with the forward modeling software (\ci{Bergamini et al. in preparation}) to test the accuracy of different strong lensing models (e.g., see \Fig\ref{fig:FM_sys4}, \ref{fig:FM_sys13}, \ref{fig:FM_sys16}, \ref{fig:FM_sys24}, and \ref{fig:FM_sys29}) and to study the intrinsic physical proprieties of the lensed background sources. We finally emphasize that, as it is happening for other clusters recently observed by the JWST \citep[e.g.,][]{Treu2022, Bezanson2022, Windhorst2023}, the presented lens model and its future refinements will have a fundamental role to study the lensed high-redshift sources that will soon be observed or discovered by the JWST \citep[e.g.,][]{Vanzella2022, Borsani_2022, Castellano_2022}.

\begin{acknowledgements}
 
Based on observations collected at the European Southern Observatory for Astronomical research in the Southern Hemisphere under ESO programmes with IDs: NE-deep (15.1h): 0100.A-0763(A) (P.I. E. Vanzella), NE (2h): GTO 094.A-0115B (P.I. J. Richard), SW (11h): 094.A0525(A) (P.I. F.E. Bauer).
The Hubble Frontier Field program (HFF) is based on the data made with the NASA/ESA {\it Hubble Space Telescope}. The Space Telescope Science Institute is operated by the Association of Universities for Research in Astronomy, Inc., under NASA contract NAS 5-26555. ACS was developed under NASA Contract NAS 5-32864.
We acknowledge financial support through grants PRIN-MIUR 2015W7KAWC, 2017WSCC32, and 2020SKSTHZ. 
AA has received funding from the European Union’s Horizon 2020 research and innovation programme under the Marie Skłodowska-Curie grant agreement No 101024195 — ROSEAU.
GBC thanks the Max Planck Society for support through the Max Planck Research Group for S. H. Suyu and the academic support from the German Centre for Cosmological Lensing. MM acknowledges support from the Italian Space Agency (ASI) through contract ``Euclid - Phase D". We acknowledge funding from the INAF ``main-stream'' grants 1.05.01.86.20 and  1.05.01.86.31.
\end{acknowledgements}

%
%

\bibliographystyle{aa}
\bibliography{bibliography}
 
\begin{appendix}

\section{Multiple images}
In \T\ref{tab:multiple_images}, we show the catalog of the 237 secure multiple images used as constraints in our new high-precision strong lensing model of the galaxy cluster \CL. This currently represents the largest dataset of secure multiple images used in a lens model.
\vspace{0.08cm}
\begin{table}[h]
\caption{Catalog of the spectroscopic multiple images included in the SL modeling of \CL.
} 
\label{tab:multiple_images}

\centering
\begin{tabular}{cccccc}
\hline\hline
ID & R.A. & Decl & $\rm z_{spec}$\\ 
 & deg & deg &\\ 
\hline
\vspace{-0.3cm}\\

12.1b & 64.036838 & -24.067456 & 0.940\\ 
12.1c & 64.036504 & -24.067024 & 0.940\\ 
12.2b\tablefootmark{b} & 64.036658 & -24.067316 & 0.940\\ 
12.2c\tablefootmark{b} & 64.036592 & -24.067231 & 0.940\\ 
12.3b\tablefootmark{b} & 64.036567 & -24.067368 & 0.940\\ 
12.3c\tablefootmark{b} & 64.036496 & -24.067272 & 0.940\\ 
12.4b\tablefootmark{b} & 64.036283 & -24.067485 & 0.940\\ 
12.4c\tablefootmark{b} & 64.036267 & -24.067462 & 0.940\\ 
12.5b\tablefootmark{b} & 64.036904 & -24.067201 & 0.940\\ 
12.5c\tablefootmark{b} & 64.036833 & -24.067101 & 0.940\\ 
12.6b\tablefootmark{b} & 64.036608 & -24.067572 & 0.940\\ 
12.6c\tablefootmark{b} & 64.036292 & -24.067157 & 0.940\\ 
13a & 64.039245 & -24.070383 & 1.005\\ 
13b & 64.038301 & -24.069728 & 1.005\\ 
13c & 64.034234 & -24.066016 & 1.005\\ 
201a\tablefootmark{a} & 64.040364 & -24.073056 & 1.147\\ 
201c\tablefootmark{a} & 64.033270 & -24.067470 & 1.147\\ 
24a & 64.035833 & -24.081321 & 1.637\\ 
24b & 64.031039 & -24.078953 & 1.637\\ 
24c & 64.026239 & -24.074337 & 1.637\\ 
5.1a & 64.047367 & -24.068671 & 1.893\\ 
5.1b & 64.043479 & -24.063523 & 1.893\\ 
5.1c & 64.040783 & -24.061609 & 1.893\\ 
5.2a\tablefootmark{b} & 64.047462 & -24.068823 & 1.893\\ 
5.2b\tablefootmark{b} & 64.043075 & -24.063084 & 1.893\\ 
5.2c\tablefootmark{b} & 64.041083 & -24.061802 & 1.893\\ 
5.3a & 64.047483 & -24.068851 & 1.893\\ 
5.3b & 64.043021 & -24.063021 & 1.893\\ 
5.3c & 64.041162 & -24.061848 & 1.893\\ 
5.4a\tablefootmark{b} & 64.047583 & -24.068884 & 1.893\\ 
5.4b\tablefootmark{b} & 64.042908 & -24.062865 & 1.893\\ 
5.4c\tablefootmark{b} & 64.041479 & -24.061979 & 1.893\\ 
5.5a\tablefootmark{b} & 64.047650 & -24.068971 & 1.893\\ 
5.5b\tablefootmark{b} & 64.042762 & -24.062771 & 1.893\\ 
5.5c\tablefootmark{b} & 64.041704 & -24.062128 & 1.893\\ 
5.6a\tablefootmark{b} & 64.047737 & -24.069012 & 1.893\\ 
5.6b\tablefootmark{b} & 64.042571 & -24.062628 & 1.893\\ 
5.6c\tablefootmark{b} & 64.042071 & -24.062319 & 1.893\\ 
36a & 64.031620 & -24.085769 & 1.964\\ 
36b & 64.028335 & -24.084550 & 1.964\\

\hline  
\end{tabular}
\tablefoot{\\
\tablefoottext{a}{New images with respect to \citetalias{Bergamini_2021}.}\\
\tablefoottext{b}{New images not included in the \citetalias{Richard_2021} lens catalog.}\\
\tablefoottext{c}{Images in common with \citetalias{Richard_2021} for which we revisited the redshift measurement.}\\
\tablefoottext{d}{Images of Sys16 with corrected positions and associations with respect to \citetalias{Bergamini_2021} (see \Sec\ref{sec:multiple_images} and \Fig\ref{fig:FM_sys16}).}
}
\end{table}

\begin{table}[h]

\centering
\begin{tabular}{cccccc}
\hline\hline
ID & R.A. & Decl & $\rm z_{spec}$\\ 
 & deg & deg & \\ 
\hline
\vspace{-0.3cm}\\

36c & 64.024074 & -24.080895 & 1.964\\ 
15.1a & 64.041804 & -24.075731 & 1.990\\ 
15.1b & 64.035250 & -24.070988 & 1.990\\ 
15.1c & 64.030771 & -24.067126 & 1.990\\ 
15.2a & 64.041833 & -24.075826 & 1.990\\ 
15.2b & 64.035171 & -24.071002 & 1.990\\ 
15.2c & 64.030771 & -24.067217 & 1.990\\ 
15.4a\tablefootmark{b} & 64.042096 & -24.075976 & 1.990\\ 
15.4b\tablefootmark{b} & 64.035008 & -24.070843 & 1.990\\ 
15.4c\tablefootmark{b} & 64.030996 & -24.067308 & 1.990\\ 
7a & 64.047102 & -24.071107 & 2.085\\ 
7b & 64.040664 & -24.063586 & 2.085\\
7c & 64.039795 & -24.063081 & 2.085\\ 
202.1b\tablefootmark{a,b} & 64.036933 & -24.066478 & 2.091\\ 
202.1c\tablefootmark{a,b} & 64.036406 & -24.066004 & 2.091\\ 
202.2b\tablefootmark{a,c} & 64.036847 & -24.066341 & 2.091\\ 
202.2c\tablefootmark{a,c} & 64.036601 & -24.066122 & 2.091\\ 
10a & 64.044564 & -24.072092 & 2.093\\ 
10b & 64.039576 & -24.066623 & 2.093\\ 
10c & 64.034336 & -24.063734 & 2.093\\ 
16.1a\tablefootmark{b} & 64.033608 & -24.069504 & 2.095\\ 
16.1b\tablefootmark{b} & 64.032596 & -24.068608 & 2.095\\ 
16.1c\tablefootmark{a,b,d} & 64.032372 & -24.068368 & 2.095\\ 
16.2a & 64.033525 & -24.069450 & 2.095\\ 
16.2b & 64.032655 & -24.068662 & 2.095\\ 
16.2c\tablefootmark{d} & 64.032415 & -24.068415 & 2.095\\ 
16.2d\tablefootmark{b,d} & 64.032450 & -24.068436 & 2.095\\ 
16.2e\tablefootmark{a,b,d} & 64.032426 & -24.068486 & 2.095\\ 
4a & 64.048126 & -24.066957 & 2.107\\ 
4b & 64.047468 & -24.066039 & 2.107\\ 
4c & 64.042209 & -24.060541 & 2.107\\ 
37a & 64.029809 & -24.086363 & 2.218\\ 
37b & 64.028610 & -24.085973 & 2.218\\ 
37c & 64.023345 & -24.081580 & 2.218\\ 
203b\tablefootmark{a} & 64.041082 & -24.063064 & 2.245\\ 
203c\tablefootmark{a} & 64.040921 & -24.062958 & 2.245\\ 
8a & 64.044624 & -24.071488 & 2.282\\ 
8b & 64.040485 & -24.066330 & 2.282\\ 
8c & 64.034256 & -24.063003 & 2.282\\ 
29a & 64.036696 & -24.083910 & 2.298\\ 
29b & 64.028408 & -24.079743 & 2.298\\ 
29c & 64.026054 & -24.077252 & 2.298\\ 
23a & 64.035668 & -24.079920 & 2.542\\ 
23b & 64.032638 & -24.078508 & 2.542\\ 
107a & 64.046032 & -24.068796 & 2.921\\ 
107b & 64.044766 & -24.066694 & 2.921\\ 
107c & 64.036203 & -24.060649 & 2.921\\ 
109a & 64.043756 & -24.073669 & 2.989\\ 
109b & 64.037760 & -24.068837 & 2.989\\ 
26a & 64.037722 & -24.082388 & 3.081\\ 
26b & 64.030484 & -24.079222 & 3.081\\ 
26c & 64.025186 & -24.073575 & 3.081\\ 
25a & 64.038073 & -24.082404 & 3.111\\ 
25b & 64.030366 & -24.079015 & 3.111\\ 
25c & 64.025446 & -24.073648 & 3.111\\ 
14.1a & 64.034492 & -24.066956 & 3.221\\ 
14.1b & 64.034188 & -24.066485 & 3.221\\ 
14.1c & 64.034000 & -24.066439 & 3.221\\

\hline  
\end{tabular}

\end{table}
\begin{table}[h]

\centering
\begin{tabular}{cccccc}
\hline\hline
ID & R.A. & Decl & $\rm z_{spec}$\\ 
 & deg & deg & \\ 
\hline
\vspace{-0.3cm}\\

14.1d & 64.033967 & -24.066901 & 3.221\\ 
14.1e\tablefootmark{b} & 64.035171 & -24.067919 & 3.221\\ 
14.1f & 64.046059 & -24.076789 & 3.221\\ 
14.2a\tablefootmark{b} & 64.034467 & -24.066860 & 3.221\\ 
14.2b\tablefootmark{b} & 64.034309 & -24.066548 & 3.221\\ 
14.2c\tablefootmark{b} & 64.034000 & -24.066439 & 3.221\\ 
20.1a & 64.040350 & -24.081474 & 3.222\\ 
20.1b & 64.032165 & -24.075108 & 3.222\\ 
20.1c & 64.027571 & -24.072671 & 3.222\\ 
20.3a\tablefootmark{b} & 64.040325 & -24.081228 & 3.222\\ 
20.3c\tablefootmark{b} & 64.027454 & -24.072209 & 3.222\\ 
20.4a\tablefootmark{a,b} & 64.040501 & -24.081207 & 3.222\\ 
20.4c\tablefootmark{a,b} & 64.027635 & -24.072207 & 3.222\\ 
20.5a\tablefootmark{a,b} & 64.040312 & -24.081729 & 3.222\\ 
20.5b\tablefootmark{a,b} & 64.032001 & -24.075318 & 3.222\\ 
20.5c\tablefootmark{a,b} & 64.027639 & -24.073080 & 3.222\\ 
20.6a\tablefootmark{a,b} & 64.040355 & -24.081772 & 3.222\\ 
20.6b\tablefootmark{a,b} & 64.031940 & -24.075309 & 3.222\\ 
20.6c\tablefootmark{a,b} & 64.027729 & -24.073147 & 3.222\\ 
20.7a\tablefootmark{a,b} & 64.040410 & -24.081780 & 3.222\\ 
20.7b\tablefootmark{a,b} & 64.031885 & -24.075296 & 3.222\\ 
20.7c\tablefootmark{a,b} & 64.027829 & -24.073199 & 3.222\\ 
1a & 64.049084 & -24.062862 & 3.238\\ 
1b & 64.046959 & -24.060797 & 3.238\\ 
1c & 64.046449 & -24.060397 & 3.238\\ 
28a & 64.038350 & -24.084126 & 3.253\\ 
28b & 64.028322 & -24.079004 & 3.253\\ 
28c & 64.026330 & -24.076705 & 3.253\\ 
3a & 64.049233 & -24.068184 & 3.290\\ 
3b & 64.045269 & -24.062763 & 3.290\\ 
3c & 64.041561 & -24.060001 & 3.290\\ 
204a\tablefootmark{a} & 64.035669 & -24.083587 & 3.291\\ 
204b\tablefootmark{a} & 64.030101 & -24.080928 & 3.291\\ 
204c\tablefootmark{a} & 64.023845 & -24.075001 & 3.291\\ 
11a\tablefootmark{b} & 64.046841 & -24.075385 & 3.292\\ 
11b & 64.038515 & -24.065965 & 3.292\\ 
11c & 64.035227 & -24.064737 & 3.292\\ 
9.1a & 64.045111 & -24.072345 & 3.292\\ 
9.1b & 64.040079 & -24.066738 & 3.292\\ 
9.2a & 64.045356 & -24.072524 & 3.292\\ 
9.2b & 64.039996 & -24.066651 & 3.292\\ 
9.3a\tablefootmark{b} & 64.045511 & -24.072678 & 3.292\\ 
9.3b\tablefootmark{b} & 64.039925 & -24.066616 & 3.292\\ 
30a & 64.033628 & -24.083185 & 3.440\\ 
30b & 64.031251 & -24.081904 & 3.440\\ 
30c & 64.022699 & -24.074595 & 3.440\\ 
27a & 64.037469 & -24.083657 & 3.492\\ 
27b & 64.029409 & -24.079889 & 3.492\\ 
27c & 64.024946 & -24.075021 & 3.492\\ 
6a & 64.047808 & -24.070164 & 3.607\\ 
6b & 64.043657 & -24.064401 & 3.607\\ 
6c & 64.037676 & -24.060756 & 3.607\\ 
205a\tablefootmark{a} & 64.035971 & -24.082554 & 3.715\\ 
205b\tablefootmark{a} & 64.031080 & -24.080300 & 3.715\\ 
205c\tablefootmark{a,b} & 64.023563 & -24.073171 & 3.715\\ 
18a & 64.040177 & -24.079872 & 3.871\\ 
18b & 64.033937 & -24.074565 & 3.871\\ 

\hline  
\end{tabular}
\end{table}

\begin{table}[h]

\centering
\begin{tabular}{cccccc}
\hline\hline
ID & R.A. & Decl & $\rm z_{spec}$\\ 
 & deg & deg & \\ 
\hline
\vspace{-0.3cm}\\
34b & 64.027632 & -24.082609 & 3.923\\ 
34c & 64.023731 & -24.078477 & 3.923\\ 
17a & 64.040496 & -24.078397 & 3.966\\ 
17b & 64.035108 & -24.073855 & 3.966\\ 
17c & 64.027163 & -24.068238 & 3.966\\ 
104a & 64.043915 & -24.075057 & 4.070\\ 
104b & 64.037239 & -24.069682 & 4.070\\ 
105a & 64.046427 & -24.076733 & 4.071\\ 
105b & 64.035986 & -24.067871 & 4.071\\ 
105c & 64.033727 & -24.065794 & 4.071\\ 
19.1a & 64.040127 & -24.080318 & 4.103\\ 
19.1b & 64.033667 & -24.074766 & 4.103\\ 
19.1c & 64.026599 & -24.070498 & 4.103\\ 
19.2a\tablefootmark{a,b} & 64.040157 & -24.080302 & 4.103\\ 
19.2c\tablefootmark{a,b} & 64.026642 & -24.070479 & 4.103\\ 
19.3a\tablefootmark{a,b} & 64.040075 & -24.080319 & 4.103\\ 
19.3b\tablefootmark{a,b} & 64.033690 & -24.074827 & 4.103\\ 
19.3c\tablefootmark{a,b} & 64.026532 & -24.070496 & 4.103\\ 
103a & 64.048183 & -24.070892 & 4.115\\ 
103b & 64.042892 & -24.063898 & 4.115\\ 
103c & 64.037669 & -24.061026 & 4.115\\ 
106a & 64.047745 & -24.068648 & 4.116\\ 
106b & 64.045866 & -24.065809 & 4.116\\ 
106c & 64.037746 & -24.059831 & 4.116\\ 
31a & 64.035486 & -24.084679 & 4.122\\ 
31b & 64.029234 & -24.081813 & 4.122\\ 
31c & 64.023412 & -24.076125 & 4.122\\ 
110a & 64.042733 & -24.072187 & 4.298\\ 
110b & 64.039160 & -24.069769 & 4.298\\ 
101a & 64.048082 & -24.074314 & 4.299\\ 
101b & 64.039685 & -24.064269 & 4.299\\ 
101c & 64.036549 & -24.063271 & 4.299\\ 
207a\tablefootmark{a} & 64.039233 & -24.081281 & 4.502\\ 
207b\tablefootmark{a} & 64.033087 & -24.076003 & 4.502\\ 
207c\tablefootmark{a} & 64.025604 & -24.071225 & 4.502\\ 
208b\tablefootmark{a} & 64.027742 & -24.080670 & 4.530\\ 
208c\tablefootmark{a} & 64.024710 & -24.077115 & 4.530\\ 
108a & 64.046513 & -24.076163 & 4.607\\ 
108b & 64.036659 & -24.068027 & 4.607\\ 
108c & 64.033508 & -24.065017 & 4.607\\ 
209a\tablefootmark{a} & 64.049597 & -24.068114 & 5.100\\ 
209b\tablefootmark{a} & 64.046853 & -24.063874 & 5.100\\ 
209c\tablefootmark{a} & 64.040622 & -24.059194 & 5.100\\ 
21.1b\tablefootmark{a,b} & 64.030897 & -24.074186 & 5.106\\ 
21.1c\tablefootmark{a,b} & 64.029225 & -24.073233 & 5.106\\ 
21.2b & 64.030775 & -24.074169 & 5.106\\ 
21.2c & 64.029292 & -24.073327 & 5.106\\ 
21.3b\tablefootmark{a,b} & 64.030855 & -24.074329 & 5.106\\ 
21.3c\tablefootmark{a,b} & 64.029169 & -24.073386 & 5.106\\ 
32a & 64.035054 & -24.085509 & 5.365\\ 
32b & 64.028403 & -24.082993 & 5.365\\ 
32c & 64.022988 & -24.077265 & 5.365\\ 
35a & 64.033729 & -24.085702 & 5.638\\ 
35b & 64.028662 & -24.084216 & 5.638\\ 
35c & 64.022125 & -24.077279 & 5.638\\ 
33a & 64.032017 & -24.084230 & 5.973\\ 
33b & 64.030821 & -24.083697 & 5.973\\ 

\hline  
\end{tabular}
\end{table}

\begin{table}[h]

\centering
\begin{tabular}{cccccc}
\hline\hline
ID & R.A. & Decl & $\rm z_{spec}$\\ 
 & deg & deg & \\ 
\hline
\vspace{-0.3cm}\\

113a & 64.045972 & -24.074033 & 5.995\\ 
113b & 64.039850 & -24.066907 & 5.995\\ 
102a & 64.048412 & -24.073606 & 6.064\\ 
102b & 64.040998 & -24.064084 & 6.064\\ 
102c & 64.036405 & -24.062218 & 6.064\\ 
2.1a & 64.050874 & -24.066542 & 6.145\\ 
2.1b & 64.047843 & -24.062059 & 6.145\\ 
2.1c & 64.043569 & -24.059003 & 6.145\\ 
2.2a\tablefootmark{b} & 64.050804 & -24.066410 & 6.145\\ 
2.2b & 64.048175 & -24.062403 & 6.145\\ 
2.2c & 64.043408 & -24.058915 & 6.145\\ 
112a\tablefootmark{b} & 64.049288 & -24.070949 & 6.145\\ 
112b & 64.043300 & -24.062949 & 6.145\\ 
112c & 64.038892 & -24.060640 & 6.145\\ 
210.1a\tablefootmark{a,b} & 64.050692 & -24.065825 & 6.149\\ 
210.1b\tablefootmark{a,b} & 64.049236 & -24.063344 & 6.149\\ 
210.2b\tablefootmark{a,b} & 64.047214 & -24.061244 & 6.149\\ 
210.2c\tablefootmark{a,b} & 64.044422 & -24.059295 & 6.149\\ 
210.3a\tablefootmark{a,b} & 64.051078 & -24.066512 & 6.149\\ 
210.3b\tablefootmark{a,b} & 64.047126 & -24.061139 & 6.149\\ 
210.3c\tablefootmark{a,b} & 64.044490 & -24.059323 & 6.149\\ 
210.4b\tablefootmark{a,b} & 64.041495 & -24.062630 & 6.149\\ 
210.4c\tablefootmark{a,b} & 64.040105 & -24.061874 & 6.149\\ 
211b\tablefootmark{a,b} & 64.045834 & -24.060190 & 6.629\\ 
211c\tablefootmark{a,b} & 64.045535 & -24.060045 & 6.629\\ 
\hline  
\end{tabular}
\end{table}

\end{appendix}

\end{document}